\documentclass[12pt]{article}
\usepackage{amsfonts}
\usepackage{bm}
\begin{document}

\renewcommand{\theequation}{\thesection.\arabic{equation}}

\title{Whitham systems and deformations.}

\author{A.Ya. Maltsev}

\date{
\centerline{L.D.Landau Institute for Theoretical Physics,}
\centerline{119334 ul. Kosygina 2, Moscow, maltsev@itp.ac.ru}}

\maketitle

\begin{abstract}
We consider the deformations of Whitham systems
including the "dispersion terms" and having the form of
Dubrovin-Zhang deformations of Frobenius manifolds.
The procedure is connected with B.A. Dubrovin problem of
deformations of Frobenius manifolds corresponding to the
Whitham systems of integrable hierarchies.
Under some non-degeneracy requirements we suggest a
general scheme of the deformation of the hyperbolic Whitham 
systems using the initial non-linear system. The general
form of the deformed Whitham system coincides with the
form of the "low-dispersion" asymptotic expansions used 
by B.A. Dubrovin and Y. Zhang in the theory of 
deformations of Frobenius manifolds. 
\end{abstract}

\section{Introduction.}

 The classical Whitham method (\cite{whith1,whith2,whith3,luke})
is connected with the slow modulations of the exact periodic or
quasiperiodic solutions of non-linear PDE's

\begin{equation}
\label{insyst}
F^{i} (\bm{\varphi}, \bm{\varphi}_{t}, \bm{\varphi}_{x},
\bm{\varphi}_{tt}, \bm{\varphi}_{xt}, \bm{\varphi}_{xx}, \dots) 
\, = \, 0 \,\,\,\,\,\,\,\, , \,\,\,\,\, i = 1, \dots, n 
\end{equation}
where $\bm{\varphi} \, = \, (\varphi^{1}, \dots, \varphi^{n})$.

 It is assumed that the system (\ref{insyst}) admits the 
finite-parametric family of exact solutions

\begin{equation}
\label{qps}
\varphi^{i} (x,t) \, = \, \Phi^{i} 
({\bf k} ({\bf U}) x \, + \, \bm{\omega} ({\bf U}) t \, + \,
\bm{\theta}_{0}, \, {\bf U})
\end{equation}
where $\bm{\theta} \, = \, (\theta^{1}, \dots, \theta^{m})$,
$\Phi^{i} (\bm{\theta}, {\bf U})$ are smooth functions 
$2\pi$-periodic w.r.t. each $\theta^{\alpha}$,
${\bf k} ({\bf U}) \, = \, (k^{1}({\bf U}), \dots, k^{m}({\bf U}))$,
$\bm{\omega} ({\bf U}) \, = \,
(\omega^{1}({\bf U}), \dots, \omega^{m}({\bf U}))$ are
"wave numbers" and "frequencies" of the solution,
${\bf U} \, = \, (U^{1}, \dots, U^{N})$ are parameters of the
solution, and 
$\bm{\theta}_{0} \, = \, (\theta^{1}_{0}, \dots, \theta^{m}_{0})$
are arbitrary initial phases.
 
 The functions $\bm{\Phi} (\bm{\theta}, {\bf U})$ satisfy the
nonlinear system

\begin{equation}
\label{phasesyst}
F^{i} \left( \bm{\Phi}, \,
\omega^{\alpha}({\bf U}) \, \bm{\Phi}_{\theta^{\alpha}}, \, 
k^{\beta}({\bf U}) \, \bm{\Phi}_{\theta^{\beta}}, \,
\omega^{\gamma}({\bf U}) \, \omega^{\delta}({\bf U}) \, 
\bm{\Phi}_{\theta^{\gamma}\theta^{\delta}}, \, \dots \right) 
\,\,\,\,\, = \,\,\,\,\, 0
\end{equation}

 We can introduce the families $\Lambda_{{\bf k},\bm{\omega}}$
and the full family 
$\Lambda = \bigcup \Lambda_{{\bf k},\bm{\omega}}$
of the functions $\bm{\Phi} (\bm{\theta}, {\bf U})$ 
satisfying the system (\ref{phasesyst}) in the space of
$2\pi$-periodic w.r.t. each $\theta^{\alpha}$ functions.
Let us choose (in a smooth way) at every
$(U^{1}, \dots, U^{N})$
some function $\bm{\Phi}(\bm{\theta}, {\bf U})$ as
having "zero initial phase shifts" and represent the
full family of $m$-phase solutions of system (\ref{insyst}) 
in the form (\ref{qps}).

\vspace{0.5cm}

 In Whitham method we make a rescaling
$X = \epsilon \, x$, $T = \epsilon \, t$  
($\epsilon \rightarrow 0$) of both variables $x$ and
$t$ and try to find a function

\begin{equation}
\label{Sfunc}
{\bf S}(X,T) \, = \, \left(
S^{1}(X,T), \dots, S^{m}(X,T) \right)
\end{equation}
and $2\pi$-periodic functions

\begin{equation}
\label{epsexp}
\Psi^{i} (\bm{\theta},X,T,\epsilon) \,\,\, = \,\,\,
\sum_{k \geq 0} \, \Psi_{(k)}^{i} (\bm{\theta},X,T) \,\,
\epsilon^{k}
\end{equation}
such that the functions

\begin{equation}
\label{whithsol}
\phi^{i} (\bm{\theta},X,T,\epsilon) \,\,\, = \,\,\,
\Psi^{i} \left({{\bf S}(X,T) \over
\epsilon} \, + \, \bm{\theta},X,T,\epsilon \right)
\end{equation}
satisfy the system

\begin{equation}
\label{epssyst} 
F^{i} \left(\bm{\phi}, \, \epsilon \, \bm{\phi}_{T}, \,
\epsilon \, \bm{\phi}_{X}, \, \epsilon^{2} \, \bm{\phi}_{TT}, 
\, \dots \right) \,\,\, = \,\,\, 0
\end{equation}
at every $X$, $T$ and $\bm{\theta}$.

 It is easy to see that the function
$\bm{\Psi}_{(0)}(\bm{\theta},X,T)$ satisfies the
system (\ref{phasesyst}) at every $X$ and $T$ with 

$$k^{\alpha} \,\,\, = \,\,\, S^{\alpha}_{X}
\,\,\,\,\, , \,\,\,\,\,
\omega^{\alpha} \,\,\, = \,\,\, S^{\alpha}_{T} $$
and so belongs at every $(X,T)$ to the family $\Lambda$.
We can write then

$$\Psi^{i}_{(0)}(\bm{\theta},X,T) \,\,\, = \,\,\,  
\Phi^{i} (\bm{\theta} \, + \, \bm{\theta}_{0}(X,T),
{\bf U}(X,T))$$
and introduce the functions $U^{\nu}(X,T)$,
$\theta^{\alpha}_{0}(X,T)$ as the parameters characterizing
the main term in (\ref{epsexp}) which should satisfy the
condition

\begin{equation}
\label{comcond}
\left[k^{\alpha}({\bf U})\right]_{T} \,\,\, = \,\,\,
\left[\omega^{\alpha}({\bf U})\right]_{X}
\end{equation}

 The functions $\Psi_{(1)}^{i} (\bm{\theta},X,T)$ 
are defined from the linear system

\begin{equation}
\label{firstappr}
{\hat L}^{i}_{j} \,
\Psi_{(1)}^{j} (\bm{\theta},X,T) \,\,\, = \,\,\,
f^{i}_{(1)} (\bm{\theta},X,T)
\end{equation}
where

$${\hat L}^{i}_{j} \,\,\,\,\, = \,\,\,\,\,
{\hat L}^{i}_{(X,T) \, j} \,\,\,\,\, = \,\,\,\,\,
{\partial F^{i} \over \partial \varphi^{j}}
\left(\bm{\Psi}_{(0)} (\bm{\theta},X,T), \, \dots \right) 
\, + $$
\begin{equation}
\label{linsyst}
+ \, {\partial F^{i} \over \partial \varphi^{j}_{t}}
\left(\bm{\Psi}_{(0)} (\bm{\theta},X,T), \, \dots \right) \,
\omega^{\alpha}(X,T) \, {\partial \over \partial \theta^{\alpha}}
\, + \, {\partial F^{i} \over \partial \varphi^{j}_{x}}
\left(\bm{\Psi}_{(0)} (\bm{\theta},X,T), \, \dots \right) \,
k^{\beta}(X,T) \, {\partial \over \partial \theta^{\beta}} 
\, + \, \dots
\end{equation}
is the linearization of system (\ref{phasesyst}) and
${\bf f}_{(1)} (\bm{\theta},X,T)$ is discrepancy given by
the expression

$$f^{i}_{(1)} (\bm{\theta},X,T) \, = \, - \,
{\partial F^{i} \over \partial \varphi^{j}_{t}} 
\left(\bm{\Psi}_{(0)} (\bm{\theta},X,T), \, \dots \right) \,
\Psi^{j}_{(0)T} (\bm{\theta},X,T) \, - $$
$$- \, {\partial F^{i} \over \partial \varphi^{j}_{x}}
\left(\bm{\Psi}_{(0)} (\bm{\theta},X,T), \, \dots \right) \,
\Psi^{j}_{(0)X} (\bm{\theta},X,T) \, - $$
\begin{equation}
\label{descreap}
- \, {\partial F^{i} \over \partial \varphi^{j}_{tt}}
\left(\bm{\Psi}_{(0)} (\bm{\theta},X,T), \, \dots \right) \,
\left(2 \, \omega^{\alpha} (X,T) \, \Psi^{j}_{(0)\theta^{\alpha}T}
\, + \, \omega^{\beta}_{T} (X,T) \,
\Psi^{j}_{(0)\theta^{\beta}} \right) \, - \, \dots
\end{equation}

 We have here

$${\partial \over \partial T} \,\,\, = \,\,\, 
U^{\nu}_{T} \, {\partial \over \partial U^{\nu}}
\,\,\, + \,\,\,
\theta^{\alpha}_{(0)T} \,
{\partial \over \partial \theta^{\alpha}}
\,\,\,\,\, , \,\,\,\,\,
{\partial \over \partial X} \,\,\, = \,\,\,
U^{\nu}_{X} \, {\partial \over \partial U^{\nu}}
\,\,\, + \,\,\,
\theta^{\alpha}_{(0)X} \,
{\partial \over \partial \theta^{\alpha}} $$   
for the functions

$$\Psi^{i}_{(0)} (\bm{\theta},X,T) \,\,\, = \,\,\,
\Phi^{i} (\bm{\theta} \, + \, \bm{\theta}_{0}(X,T),
{\bf U}(X,T))$$

 We will assume that $k^{\alpha}$ and $\omega^{\alpha}$
can be considered (locally) as the independent parameters
on the family $\Lambda$ and the total family of solutions
of (\ref{phasesyst}) depends (for generic $k^{\alpha}$,
$\omega^{\alpha}$) on $N \, = \, 2m \, + \, s$,
($s \geq 0$) parameters $U^{\nu}$ and $m$ initial
phases $\theta_{0}^{\alpha}$.

 Easy to see that the functions
$\bm{\Phi}_{\theta^{\alpha}} (\bm{\theta}  + 
\bm{\theta}_{0}(X,T), {\bf U}(X,T))$
and \linebreak
$\nabla_{\bm{\xi}} \, \bm{\Phi}_{\theta^{\alpha}}
(\bm{\theta} \, + \, \bm{\theta}_{0}(X,T), {\bf U}(X,T))$
where $\bm{\xi}$ is any vector in the space of parameters
$U^{\nu}$ tangential to the surface
${\bf k} \, = \, const$, $\bm{\omega} \, = \, const$
belong to the kernel of the operator
${\hat L}^{i}_{(X,T) \, j}$.

\vspace {0.5cm}

 Let us put now some "regularity" conditions on the family
(\ref{qps}) of quasiperiodic solutions of (\ref{insyst})

\vspace {0.5cm}

{\bf Definition 1.1.}

{\it We call the family (\ref{qps}) the full regular   
family of $m$-phase solutions of (\ref{insyst}) if:

1) The functions
$\bm{\Phi}_{\theta^{\alpha}} (\bm{\theta}, {\bf U})$,
$\bm{\Phi}_{U^{\nu}} (\bm{\theta}, {\bf U})$ are linearly
independent (almost everywhere) on the set $\Lambda$;

2) The $m \, + \, s$ linearly independent functions
$\bm{\Phi}_{\theta^{\alpha}} (\bm{\theta}, {\bf U})$,
$\nabla_{\bm{\xi}} \, \bm{\Phi} (\bm{\theta}, {\bf U})$  
($\nabla_{\bm{\xi}} {\bf k} \, = \, 0$,
$\nabla_{\bm{\xi}} \bm{\omega} \, = \, 0$)
give the full kernel of the operator
${\hat L}^{i}_{[{\bf U}] \, j}$ 
(here $\bm{\theta}_{0} \, = \, 0$) for generic
${\bf k}$ and $\bm{\omega}$.
 
3) There are exactly $m \, + \, s$ linearly independent  
"left eigen-vectors"
$\bm{\kappa}^{(q)}_{[{\bf U}]}(\bm{\theta})$,
$q = 1, \dots, m+s$ of the operator
${\hat L}^{i}_{[{\bf U}] \, j}$ (for generic
${\bf k}$ and $\bm{\omega}$) corresponding to zero
eigen values i.e.

$$\int_{0}^{2\pi}\!\!\!\dots\int_{0}^{2\pi}
\kappa_{[{\bf U}]i}^{(q)}(\bm{\theta}) \,
{\hat L}^{i}_{[{\bf U}] \, j} \,
\psi^{j} (\bm{\theta}) \,
{d^{m} \theta \over (2\pi)^{m}} \,\,\,
\equiv \,\,\, 0$$
for any periodic $\psi^{j} (\bm{\theta})$.
}

\vspace{0.5cm}

 It's not difficult to see that the Definition 1.1 is connected
with the regularity properties of the sub-manifold $\Lambda$
given by the set of functions $\bm{\Phi} (\bm{\theta}, {\bf U})$
in the space of $2\pi$-periodic functions. In fact, for our 
purposes we can use also the weaker definition of the full regular 
family of $m$-phase solutions of (\ref{insyst}). Namely,
let us represent the space of parameters ${\bf U}$ in the form
${\bf U} = ({\bf k}, \bm{\omega}, {\bf n})$ where ${\bf k}$
are the wave numbers, $\bm{\omega}$ are the frequencies of
$m$-phase solutions, and ${\bf n} = (n^{1}, \dots, n^{s})$
are some additional parameters (if they exist). Let us give now
the "weak" definition of the full regular family of
$m$-phase solutions in the form:

\vspace{0.5cm}

{\bf Definition 1.1$^{\prime}$.}

{\it
We call the family $\Lambda$ the full regular family of
$m$-phase solutions of (\ref{insyst}) if

1) The functions
$\bm{\Phi}_{\theta^{\alpha}} (\bm{\theta},
{\bf k}, \bm{\omega}, {\bf n})$,
$\bm{\Phi}_{n^{l}} (\bm{\theta}, {\bf k}, \bm{\omega}, {\bf n})$
are linearly independent and give (for generic ${\bf k}$ and 
$\bm{\omega}$) the full basis in the kernel of the operator
${\hat L}^{i}_{j[\bm{\theta}_{0},{\bf k},\bm{\omega},{\bf n}]}$;
   
2) The operator
${\hat L}^{i}_{j[\bm{\theta}_{0},{\bf k},\bm{\omega},{\bf n}]}$
has (for generic ${\bf k}$ and $\bm{\omega}$) exactly $m + s$
linearly independent "left eigen vectors"

$$\bm{\kappa}^{(q)}_{[{\bf U}]} (\bm{\theta} + \bm{\theta}_{0}) 
\,\,\, = \,\,\,
\bm{\kappa}^{(q)}_{[{\bf k}, \bm{\omega}, {\bf n}]}
(\bm{\theta} + \bm{\theta}_{0}) $$
depending on the parameters ${\bf U}$ in a smooth way and
corresponding to zero eigen-values.
}

\vspace{0.5cm}

 We will assume now that the system (\ref{insyst}) has a full
regular family of quasiperiodic $m$-phase solutions in strong 
or weak sense in all our considerations.

\vspace{0.5cm}

 To find the function 
$\bm{\Psi}_{(1)} (\bm{\theta},X,T)$ we have to put now
the $m \, + \, s$ conditions of orthogonality of the discrepancy
${\bf f}_{(1)}(\bm{\theta},X,T)$ to the functions
$\bm{\kappa}^{(q)}_{[{\bf U}(X,T)]}
(\bm{\theta}\, + \, \bm{\theta}_{0}(X,T))$

\begin{equation}
\label{ortcond}
\int_{0}^{2\pi}\!\!\!\dots\int_{0}^{2\pi}   
\kappa^{(q)}_{[{\bf U}(X,T)]\, i}
(\bm{\theta}\, + \, \bm{\theta}_{0}(X,T)) \,
f^{i}_{(1)} (\bm{\theta},X,T) \,
{d^{m} \theta \over (2\pi)^{m}} \,\,\, = \,\,\, 0
\end{equation}

 The system (\ref{ortcond}) together with (\ref{comcond})
gives
$m \, + \, (m \, + \, s) \, = \, 2m \, + \, s \, = \, N$
conditions at each $X$ and $T$ on the parameters of zero
approximation $\bm{\Psi}_{(0)} (\bm{\theta},X,T)$
necessary for the construction of the first
$\epsilon$-term in the solution (\ref{epsexp}).

Let us prove now the following Lemma about the orthogonality
conditions (\ref{ortcond}):

\vspace{0.5cm}

{\bf Lemma 1.1.}  

{\it Under all the assumptions of regularity formulated above
the orthogonality conditions (\ref{ortcond}) do not contain
the functions $\theta^{\alpha}_{0}(X,T)$ and give just the
restrictions on the functions $U^{\nu}(X,T)$ having the form

\begin{equation}
\label{1whithsyst}
C^{(q)}_{\nu} ({\bf U}) \, U^{\nu}_{T} \,\, - \,\,
D^{(q)}_{\nu} ({\bf U}) \, U^{\nu}_{X} \,\, = \,\, 0 
\end{equation}
(with some functions $C^{(q)}_{\nu} ({\bf U})$,  
$D^{(q)}_{\nu} ({\bf U})$).
}
 
\vspace{0.5cm}

 Proof.

 Let us write down the part 
${\bf f}^{\prime}_{(1)}$ of the function ${\bf f}_{(1)}$
which contains the derivatives $\theta^{\beta}_{0T}(X,T)$
and $\theta^{\beta}_{0X}(X,T)$. We have from (\ref{descreap})

$$f^{\prime i}_{(1)}(\bm{\theta},X,T) \, = \, - \,
{\partial F^{i} \over \partial \varphi^{j}_{t}}
\left(\bm{\Psi}_{(0)}, \, \dots \right) \,
\Psi^{j}_{(0)\theta^{\beta}} \, \theta^{\beta}_{0T} \, - \,
{\partial F^{i} \over \partial \varphi^{j}_{x}}
\left(\bm{\Psi}_{(0)}, \, \dots \right) \,
\Psi^{j}_{(0)\theta^{\beta}} \, \theta^{\beta}_{0X} \, - $$
$$- \, {\partial F^{i} \over \partial \varphi^{j}_{tt}}
\left(\bm{\Psi}_{(0)}, \, \dots \right) \, 2 \, 
\omega^{\alpha} (X,T) \, 
\Psi^{j}_{(0)\theta^{\alpha}\theta^{\beta}} \,
\theta^{\beta}_{0T} \, - \,
\, {\partial F^{i} \over \partial \varphi^{j}_{xx}}
\left(\bm{\Psi}_{(0)}, \, \dots \right) \, 2 \,
k^{\alpha} (X,T) \,
\Psi^{j}_{(0)\theta^{\alpha}\theta^{\beta}} \,
\theta^{\beta}_{0X} \, - \, \dots $$

 Let us choose again the set of parameters ${\bf U}$ in the form

$${\bf U} = (k^{1}, \dots, k^{m}, \omega^{1}, \dots, \omega^{m},
n^{1}, \dots, n^{s})$$ 
where $k^{\alpha}$ are the wave numbers,
$\omega^{\alpha}$ are the frequencies of $m$-phase solutions
and $(n^{1}, \dots, n^{s})$ are additional parameters
(except the initial phases).

 We can write then

$$f^{\prime i}_{(1)}(\bm{\theta},X,T) \, = \, \left[ 
- \, {\partial \over \partial \omega^{\beta}} \,
F^{i} \left( \bm{\Phi} (\bm{\theta} + \bm{\theta}_{0}, {\bf U}),
\dots \right) \, + \, {\hat L}^{i}_{j} \,
{\partial \over \partial \omega^{\beta}} \, 
\Phi^{j} (\bm{\theta} + \bm{\theta}_{0}, {\bf U}) \right] \,
\theta^{\beta}_{0T} \, + \, $$
$$+ \, \left[ - \, {\partial \over \partial k^{\beta}} \,
F^{i} \left( \bm{\Phi} (\bm{\theta} + \bm{\theta}_{0}, {\bf U}),
\dots \right) \, + \, {\hat L}^{i}_{j} \,
{\partial \over \partial k^{\beta}} \,
\Phi^{j} (\bm{\theta} + \bm{\theta}_{0}, {\bf U}) \right] \,
\theta^{\beta}_{0X} $$

 The derivatives $\partial F^{i} / \partial \omega^{\beta}$ 
and $\partial F^{i} / \partial k^{\beta}$ are identically zero
on $\Lambda$ according to (\ref{phasesyst}). We have then

$$\int_{0}^{2\pi}\!\!\!\dots\int_{0}^{2\pi}
\kappa^{(q)}_{[{\bf U}(X,T)]\, i}
(\bm{\theta}\, + \, \bm{\theta}_{0}(X,T)) \,
f^{\prime i}_{(1)} (\bm{\theta},X,T) \,
{d^{m} \theta \over (2\pi)^{m}} \,\,\, \equiv \,\,\, 0 $$
since all $\bm{\kappa}^{(q)} (\bm{\theta},X,T)$ are the
left eigen-vectors of ${\hat L}$ with zero eigen-values.

 It is easy to see also that all $\bm{\theta}_{0}(X,T)$
in the arguments of $\bm{\Phi}$ and $\bm{\kappa}^{(q)}$     
will disappear after the integration w.r.t. $\bm{\theta}$
so we get the statement of the Lemma.

{\hfill Lemma 1.1 is proved.}

\vspace{0.5cm}

 The system

$${\partial k^{\alpha} \over \partial U^{\nu}} \,
U^{\nu}_{T} \,\, = \,\,
{\partial \omega^{\alpha} \over \partial U^{\nu}} \,
U^{\nu}_{X} \,\,\, , \,\,\,\,\, \alpha \, = \, 1, \dots, m $$
\begin{equation}
\label{wsyst1}
C^{(q)}_{\nu} ({\bf U}) \, U^{\nu}_{T} \,\, = \,\,
D^{(q)}_{\nu} ({\bf U}) \, U^{\nu}_{X} \,\,\, , \,\,\,\,\,
q \, = \, 1, \dots, m+s
\end{equation}
is called the Whitham system for the $m$-phase solutions of
system (\ref{insyst}).

 Let us note that we have 
$rank ||\partial k^{\alpha}/\partial U^{\nu}|| = m$
according to our assumption above. In the generic case the
derivatives $U^{\nu}_{T}$ can be expressed through $U^{\mu}_{X}$
and the Whitham system (\ref{wsyst1}) can be written in the form

\begin{equation}
\label{HTsyst}
U^{\nu}_{T} \, = \, V^{\nu}_{\mu} ({\bf U}) \, U^{\mu}_{X}
\,\,\,\,\, , \,\,\,\,\, \nu, \, \mu \, = \, 1, \dots, N
\end{equation}
where $V^{\nu}_{\mu} (U)$ is some $N \times N$ matrix
depending on the variables $U^{1}, \dots, U^{N}$.

 Let us say that quite often the system (\ref{insyst}) can
be written in the evolution form

\begin{equation}
\label{evsyst}
\varphi^{i}_{t} \,\,\, = \,\,\,
Q^{i} (\bm{\varphi}, \, \bm{\varphi}_{x}, \, \bm{\varphi}_{xx},
\, \dots \, )
\end{equation}

 For systems (\ref{evsyst}) the form (\ref{HTsyst}) of the
corresponding Whitham system has then a natural motivation.

\vspace{0.5cm}

 We will assume here that if the conditions (\ref{wsyst1})
are satisfied then the system (\ref{firstappr}) is resolvable
on the space of $2\pi$-periodic 
w.r.t. each $\theta^{\alpha}$ functions. The solution 
$\bm{\Psi}_{(1)}(\bm{\theta},X,T)$ is defined then modulo a
linear combination of the "right eigen-functions" of
${\hat L}^{i}_{(X,T)j}$ 
($\bm{\Psi}_{(0)\theta^{\alpha}}$,  $\bm{\Psi}_{(0)n^{l}}$)
introduced above. According to common approach 
(\cite{luke,dm}) we can try to use the corresponding coefficients
to make the systems analogous to (\ref{firstappr}) resolvable in the
next orders and try to find recursively all the terms of series
(\ref{epsexp}).

 Different aspects and numerous applications of the Whitham method
were studied in many different works 
(\cite{whith1}-\cite{AbGr})\footnote{We apologize for the 
impossibility to give here the full list of numerous works
on Whitham method.} 
and the Whitham method is considered now as one of the classical
methods of investigation of non-linear systems.

\vspace{0.5cm}

 It was pointed out by G. Whitham (\cite{whith1,whith2,whith3})
that the Whitham system (\ref{wsyst1}) has a local Lagrangian
structure in case when the initial system has a local Lagrangian
structure

$$\delta \,\, \int \int 
{\cal L} (\bm{\varphi}, \, \bm{\varphi}_{t}, \, \bm{\varphi}_{x} ,
\, \dots ) \,\, dx \, dt \,\,\, = \,\,\, 0 $$
on the space $\{\bm{\varphi}(x,t)\}$.

 The procedure of construction of Lagrangian formalism for the
Whitham system (\ref{wsyst1}) is given by the averaging of
the Lagrangian function ${\cal L}$ on the family of $m$-phase
solutions of system (\ref{insyst}) (\cite{whith1,whith2,whith3}).
Let us note also that in the case of presence of additional 
parameters $n^{l}$ the additional method of Whitham 
pseudo-phases should be used.

 The important procedure of averaging of local field-theoretical
Hamiltonian structures was suggested by B.A. Dubrovin and 
S.P. Novikov (\cite{dn1,Nov,dn2,dn3}). The Dubrovin-Novikov procedure
gives the local field-theoretical Hamiltonian formalism for the
Whitham system (\ref{HTsyst}) in the case when the initial system
(\ref{evsyst}) has a local Hamiltonian formalism of general type.
The Dubrovin-Novikov bracket for the Whitham system has
a general form

\begin{equation}
\label{DNbracket}
\{U^{\nu}(X), \, U^{\mu}(Y) \} \,\,\, = \,\,\,
g^{\nu\mu}({\bf U}) \,\, \delta^{\prime}(X-Y) \,\, + \,\,
b^{\nu\mu}_{\lambda} ({\bf U}) \, U^{\lambda}_{X} \,\,
\delta (X-Y) 
\end{equation}
and was called the local Poisson bracket of Hydrodynamic type.
The theory of the brackets (\ref{DNbracket}) is closely related 
with differential geometry (\cite{dn1,dn2,dn3}) and is connected
with different coordinate systems in the (pseudo) Euclidean
spaces. Let us say also that during the last years the 
important weakly-nonlocal generalizations of Dubrovin-Novikov 
brackets (Mokhov-Ferapontov bracket and Ferapontov brackets)
were introduced and studied 
(\cite{mohfer1,fer1,fer2,fer3,fer4,pavlov3,PhysD}). 

 The Hamiltonian structure (\ref{DNbracket}) for the systems 
(\ref{HTsyst}) has
a direct relation to the integrability of the systems of this 
class. Thus it was conjectured by S.P. Novikov that any 
diagonalizable system (\ref{HTsyst}) Hamiltonian with respect
to the bracket of Hydrodynamic Type is integrable. The conjecture
of S.P. Novikov was proved by S.P. Tsarev (\cite{Tsarev})
who suggested the "generalized Hodograph method" for solving
the diagonal Hamiltonian systems (\ref{HTsyst}). Let us say
that the Tsarev method has become especially important for
the Whitham systems corresponding to integrable hierarchies
and provided a lot of very important solutions for such 
systems in different cases. We note here, that the Whitham
systems of integrable hierarchies can usually be written in
diagonal form (\cite{whith3,ffm,krichev1}) and admit the 
(multi-) Hamiltonian structures given by the averaging of the 
Lagrangian or the field-theoretical Hamiltonian structures of
the initial system.

\vspace{0.5cm}

 During the last years the theory of compatible Poisson
brackets (\ref{DNbracket}) and their deformations in connection
with Quantum Field Theory was intensively developed
(\cite{Dubrov1,Dubrov2,Dubrov3,Dubrov4,DubrZhang1,Dubrov5,
DubrZhang2,Lorenzoni,DubrZhang3,LiuZhang1,LiuZhang2,DubrLiuZhang,
DubrZhangZuo}). Namely, the theory of compatible Poisson 
brackets (\ref{DNbracket}) plays the main role in the theory
of "Frobenius manifolds" constructed by B.A. Dubrovin and
connected with the classification of the Topological Quantum Field
Theories (based on the Witten - Dijkgraaf - Verlinde - Verlinde 
equation). Every "Frobenius manifold" is connected also with
the integrable hierarchy of Hydrodynamic Type

\begin{equation}
\label{HThier}
U^{\nu}_{t_{s}} \,\,\, = \,\,\,
V^{\nu}_{(s)\mu} ({\bf U}) \, U^{\mu}_{X}
\end{equation}
having the bihamiltonian structure in Dubrovin-Novikov sense
with a pair of Poisson brackets (\ref{DNbracket}).
\footnote{The hierarchy (\ref{HThier}) and the corresponding
pair of Poisson brackets possess also some additional properties
(the existence of "unit vector field", Euler field, 
$\tau$-symmetry etc. ...).}

 The " $\epsilon$-deformations " of the integrable hierarchies
and the Poisson brackets (\ref{DNbracket}) are connected with
the "higher genus" corrections in the Quantum Field Theories
and are being intensively investigated now 
(\cite{DubrZhang1,Dubrov5,DubrZhang2,Lorenzoni,DubrZhang3,LiuZhang1,
LiuZhang2,DubrLiuZhang,DubrZhangZuo}). According to B.A. Dubrovin
and Y. Zhang the $\epsilon$-deformation of Frobenius manifold
is given by the infinite series polynomial w.r.t. derivatives
${\bf U}_{X}$, ${\bf U}_{XX}$, ${\bf U}_{XXX}$, $\dots$ and
representing the "small dispersion" deformation of the hierarchy 
(\ref{HThier}) and the corresponding bihamiltonian structure of 
Hydrodynamic Type. The deformed hierarchy and bihamiltonian 
structure should possess also the additional properties of
Frobenius manifolds ("unit vector field", Euler field,
$\tau$-symmetry etc. ...) and give then the "deformation"
of the corresponding Frobenius manifold 
(\cite{DubrZhang1,DubrZhang2,DubrZhang3}). The general
problem of classification of deformations of bihamiltonian
hierarchies (\ref{HThier}) is being also intensively studied 
by now and very important results were obtained recently 
in this area (\cite{Lorenzoni,LiuZhang1,LiuZhang2,
DubrLiuZhang,DubrZhangZuo}).

 The dispersive corrections to the Whitham systems were first
considered by M.Y. Ablowitz and D.J. Benney (\cite{AblBenny},
also \cite{Abl1}-\cite{Abl2}) where the first consideration 
of multi-phase Whitham method was also made. As was shown in 
\cite{AblBenny} the dispersive corrections to Whitham systems 
can naturally arise and, besides that, can be generalized to
the multi-phase situation.

 In this paper we will consider the deformations of the Whitham
systems (\ref{HTsyst}) having Dubrovin-Zhang form, i.e. the
dispersive corrections to (\ref{HTsyst}) containing the higher 
$X$-derivatives of the parameters ${\bf U}$. The problem in this 
form is connected with the problem set by B.A. Dubrovin which
is formulated as the problem of deformations of Frobenius
manifolds corresponding to the Whitham systems of integrable
hierarchies. B.A. Dubrovin problem contains both the problem
of deformation of Whitham systems (\ref{HTsyst}) and the 
corresponding bi-Hamiltonian structures of Hydrodynamic Type
giving the dispersive corrections to these structures.

 We will consider here the first part of B.A. Dubrovin problem
and suggest a general scheme of recursive construction of terms 
of the deformation of the Whitham system (\ref{HTsyst}) using the 
initial system (\ref{insyst}) or (\ref{evsyst}). The second
part of B.A. Dubrovin problem will then be considered in the 
next paper. We will not require, however, 
the "integrability" of the system (\ref{HTsyst})
and all the considerations below can be applicable for both
integrable and non-integrable systems (\ref{insyst}),
(\ref{evsyst}). Thus, we do not require also the bihamiltonian
property of the Whitham system (\ref{HTsyst}).

In the next chapter we will make more detailed
investigation of the asymptotic series (\ref{epsexp}) and 
describe the construction of the deformation procedure for 
general Whitham systems.

\section{The deformation of the Whitham systems.}
\setcounter{equation}{0}

 Let us start again with the description of the construction
of asymptotic series (\ref{epsexp}) connected with the Whitham
method. We will assume that we have the initial system having
the general form (\ref{insyst}) which has an $N$-parametric 
(modulo the initial phases $\theta^{\alpha}_{0}$) family of
$m$-phase solutions

\begin{equation}
\label{qps1}
\varphi^{i} (x,t) \, = \, \Phi^{i}
({\bf k} ({\bf U}) x \, + \, \bm{\omega} ({\bf U}) t \, + \,
\bm{\theta}_{0}, \, {\bf U})
\end{equation}
with the functions $\bm{\Phi}(\bm{\theta}, \, {\bf U})$
satisfying the system 

\begin{equation}
\label{phsyst}
F^{i} \left( \bm{\Phi}, \,
\omega^{\alpha}({\bf U}) \, \bm{\Phi}_{\theta^{\alpha}}, \,
k^{\beta}({\bf U}) \, \bm{\Phi}_{\theta^{\beta}}, \,
\omega^{\gamma}({\bf U}) \, \omega^{\delta}({\bf U}) \,
\bm{\Phi}_{\theta^{\gamma}\theta^{\delta}}, \, \dots \right)
\,\,\,\,\, = \,\,\,\,\, 0
\end{equation}
(and $2\pi$-periodic w.r.t. all $\theta^{\alpha}$).

 The choice of the functions $\Phi^{i}(\bm{\theta}, \, {\bf U})$
at each ${\bf U}$ is defined modulo the initial phase 
$\bm{\theta}_{0}$ and the full family of solutions of 
(\ref{phsyst}) is given by all the functions 
$\bm{\Phi}(\bm{\theta} + \bm{\theta}_{0}, \, {\bf U})$
with arbitrary 
$\bm{\theta}_{0} \, = \, (\theta^{1}_{0}, \dots , \theta^{m}_{0})$.
We will just assume here that the choice of 
$\Phi^{i}(\bm{\theta}, \, {\bf U})$ is smooth on the family of
$m$-phase solutions $\Lambda$.

 We will assume also that the parameters $({\bf k}, \, \bm{\omega})$
are independent on the family $\Lambda$ such that $N \geq 2m$. It
will be convenient to use the parameters 
$(k^{1}, \dots , k^{m}, \omega^{1}, \dots , \omega^{m}, 
n^{1}, \dots , n^{s})$ on the family $\Lambda$ where $k^{\alpha}$
are the "wave numbers", $\omega^{\alpha}$ are frequencies, and
$(n^{1}, \dots , n^{s})$ are additional parameters (if they present).
The linearly independent solutions of the linearized system
(\ref{phsyst}) will then be given by $m+s$ functions
$\bm{\Phi}_{\theta^{\alpha}}(\bm{\theta}, \, {\bf U})$, \,
$\bm{\Phi}_{n^{l}}(\bm{\theta}, \, {\bf U})$.

 As we already said above we assume that the family $\Lambda$
represents the "full regular family" of $m$-phase solutions
of (\ref{insyst1}) such that all the requirements (1)-(3)
formulated in the Definition 1.1 are satisfied. We have thus
exactly $m+s$ functions 
$\bm{\Phi}_{\theta^{\alpha}}(\bm{\theta}, \, {\bf U})$, \,
$\bm{\Phi}_{n^{l}}(\bm{\theta}, \, {\bf U})$ giving the basis
in the kernel of linearized system (\ref{phsyst}) and
exactly $m+s$ linearly independent functions 
$\bm{\kappa}^{(q)}_{[{\bf U}]} (\bm{\theta} + \bm{\theta}_{0})$
giving the basis in the kernel of the adjoint operator
(for generic ${\bf k}$, $\bm{\omega}$) and depending in a smooth
way on the parameters $({\bf k},\bm{\omega},{\bf n})$. 

 As we mentioned already we make a rescaling of coordinates
$X \, = \, \epsilon \, x$,  $T \, = \, \epsilon \, t$ and try
to find the solutions of system

\begin{equation}
\label{insyst1}
F^{i} (\bm{\varphi}, \, \epsilon \, \bm{\varphi}_{T}, \,
\epsilon \, \bm{\varphi}_{X}, \,   
\epsilon^{2} \, \bm{\varphi}_{TT}, 
\epsilon^{2} \, \bm{\varphi}_{XT}, 
\epsilon^{2} \, \bm{\varphi}_{XX}, \dots)
\, = \, 0 \,\,\,\,\,\,\,\, , \,\,\,\,\, i = 1, \dots, n
\end{equation}
having the form

\begin{equation}
\label{sol1}
\phi^{i} (\bm{\theta},X,T,\epsilon) \,\,\, = \,\,\,
\Psi^{i} \left({{\bf S}(X,T) \over
\epsilon} \, + \, \bm{\theta},X,T,\epsilon \right)
\end{equation}

\begin{equation}
\label{ser1}
\Psi^{i} (\bm{\theta},X,T,\epsilon) \,\,\, = \,\,\,
\sum_{k \geq 0} \, \Psi_{(k)}^{i} (\bm{\theta},X,T) \,\,
\epsilon^{k}
\end{equation}

 The function $\bm{\Psi}_{(0)} (\bm{\theta},X,T)$ belongs to the
family $\Lambda$ at every fixed $X$ and $T$ and the compatibility
conditions (\ref{ortcond}) for the system 

\begin{equation}
\label{1syst}
{\hat L}^{i}_{j} \,
\Psi_{(1)}^{j} (\bm{\theta},X,T) \,\,\, = \,\,\,
f^{i}_{(1)} (\bm{\theta},X,T)
\end{equation}
give the Whitham system (\ref{1whithsyst}) on the parameters
${\bf U}(X,T)$ of the zero approximation.  The function 
$\bm{\Psi}_{(1)} (\bm{\theta},X,T)$ is defined from the 
system (\ref{1syst}) modulo the linear combination of functions
$\bm{\Phi}_{\theta^{\alpha}}(\bm{\theta}, \, {\bf U})$, \,
$\bm{\Phi}_{n^{l}}(\bm{\theta}, \, {\bf U})$ at every 
$X$ and $T$.

 All the other approximations $\bm{\Psi}_{(k)} (\bm{\theta},X,T)$
satisfy the linear systems

\begin{equation}
\label{ksyst}  
{\hat L}^{i}_{j} \,
\Psi_{(k)}^{j} (\bm{\theta},X,T) \,\,\, = \,\,\,
f^{i}_{(k)} (\bm{\theta},X,T)
\end{equation}
where the functions ${\bf f}_{(k)} (\bm{\theta},X,T)$
represent the higher order discrepancies given by system
(\ref{insyst1}). 

 The compatibility conditions of the systems (\ref{ksyst})
($k \geq 2$) give the restrictions on the "initial phases"
$\theta_{0}^{\alpha}$ of the zero approximation and on the
previous corrections 
$\bm{\Psi}_{(k^{\prime})} (\bm{\theta},X,T)$.

 Let us make now one more general assumption about the systems   
(\ref{1syst}) and (\ref{ksyst}). Namely we omit here the special
investigation of the solvability of systems (\ref{1syst}),
(\ref{ksyst}) on the space of $2\pi$-periodic functions
and assume that the orthogonality conditions

\begin{equation}
\label{ortcond1} 
\int_{0}^{2\pi}\!\!\!\dots\int_{0}^{2\pi}
\kappa^{(q)}_{[{\bf U}(X,T)]\, i}
(\bm{\theta}\, + \, \bm{\theta}_{0}(X,T)) \,\,\,
f^{i}_{(k)} (\bm{\theta},X,T) \,\,\,
{d^{m} \theta \over (2\pi)^{m}} \,\,\, = \,\,\, 0
\end{equation}
give the necessary and sufficient conditions of solvability
of these systems.
So, we will assume that under the conditions (\ref{ortcond1})
we can always find a smooth $2\pi$-periodic solution
of (\ref{1syst}), (\ref{ksyst}) defined modulo the linear
combination

\begin{equation}
\label{cdcomb}
\sum_{\alpha=1}^{m} \, c^{\alpha}_{(k)}(X,T) \,\,\,
\bm{\Phi}_{\theta^{\alpha}}(\bm{\theta}, \, X, \, T) 
\,\,\, + \,\,\, \sum_{l=1}^{s} \, d^{l}_{(k)}(X,T) \,\,\,
\bm{\Phi}_{n^{l}}(\bm{\theta}, \, X, \, T)
\end{equation}

\vspace{0.5cm}

 Let us consider now the construction of the asymptotic series
(\ref{ser1}). We note first that the solution 
$\bm{\Psi}_{(k)} (\bm{\theta},X,T)$ is defined from the system 
(\ref{ksyst}) modulo the linear combination (\ref{cdcomb}) which
is equivalent in the main order to the addition of the values
$\epsilon^{k+1} \, c^{\alpha}_{(k)}(X,T)$ to the phases
$S^{\alpha}(X,T)$ of the zero approximation 
$\bm{\Psi}_{(0)} (\bm{\theta},X,T)$
and to the addition of the values 
$\epsilon^{k} \, d^{l}_{(k)}(X,T)$
to the parameters $n^{l}(X,T)$ of 
$\bm{\Psi}_{(0)} (\bm{\theta},X,T)$ according to the formulae
(\ref{sol1}), (\ref{ser1}). Easy to see also that we can add the 
values $\epsilon^{k+1} \, c^{\alpha}_{(k)X}(X,T)$ and
$\epsilon^{k+1} \, c^{\alpha}_{(k)T}(X,T)$ to the parameters
$k^{\alpha}$ and $\omega^{\alpha}$ in the ${\bf U}$-dependence
of $\bm{\Psi}_{(0)} (\bm{\theta},X,T)$ which does not affect
the $k$-th order of $\epsilon$ in the series (\ref{ser1}).

\vspace{0.5cm}

 Let us change now the procedure of construction of series
(\ref{ser1}) in the following way:

\vspace{0.5cm}

1) At every step $k$ we choose the solution 
$\bm{\Psi}_{(k)} (\bm{\theta},X,T)$ in arbitrary way;

\vspace{0.5cm}

2) We allow the regular $\epsilon$-dependence 

$$S^{\alpha}(X,T,\epsilon) \,\,\, = \,\,\,
\sum_{k\geq0} \, S_{(k)}^{\alpha}(X,T) \,\, \epsilon^{k} $$

$$n^{l} (X,T,\epsilon) \,\,\, = \,\,\,
\sum_{k\geq0} \, n_{(k)}^{l} (X,T) \,\, \epsilon^{k} $$
of the phases $S^{\alpha}$ and the parameters 
$({\bf k}, \bm{\omega}, {\bf n})$ of the zero approximation
$\bm{\Psi}_{(0)} (\bm{\theta},X,T)$ such that

$$k^{\alpha} (X,T,\epsilon) \,\,\, = \,\,\,
S^{\alpha}_{X}(X,T,\epsilon) \,\,\,\,\, , \,\,\,\,\,
\omega^{\alpha} (X,T,\epsilon) \,\,\, = \,\,\,
S^{\alpha}_{T}(X,T,\epsilon) $$

\vspace{0.5cm}

3) We use the higher orders  
$S_{(k)}^{\alpha}$, $n_{(k)}^{l}$
of the functions $S^{\alpha}(X,T,\epsilon)$,
$n^{l} (X,T,\epsilon)$ to provide the orthogonality conditions
(\ref{ortcond1}) for the systems (\ref{ksyst}).

\vspace{0.5cm}

 We can write now the asymptotic solution of (\ref{insyst1})
in the form

$$\phi^{i} (\bm{\theta},X,T,\epsilon) \,\,\, = \,\,\,
\Phi^{i} \left( {{\bf S}(X,T,\epsilon) \over \epsilon} 
\, + \, \bm{\theta}, \,\,\, {\bf S}_{X}(X,T,\epsilon), \,\,\,
{\bf S}_{T}(X,T,\epsilon), \,\,\, {\bf n} (X,T,\epsilon) \right)
\,\,\, + $$
\begin{equation}
\label{ser2}
+ \,\,\, \sum_{k \geq 1} \,
{\tilde \Psi}_{(k)}^{i} \left( 
{{\bf S}(X,T,\epsilon) \over \epsilon} \, + \, \bm{\theta}, \,
X, \, T \right) \,\,\, \epsilon^{k}
\end{equation}

 Let us note that the phase shift $\bm{\theta}_{0}(X,T)$
of the initial approximation becomes now the $\epsilon$-term in 
$\epsilon$-expansion of the function ${\bf S}(X,T,\epsilon)$.

\vspace{0.5cm}

 It's not difficult to see that the series (\ref{ser2}) can
be always represented in the form (\ref{sol1})-(\ref{ser1})
and give the same set of the asymptotic solutions of system
(\ref{insyst1}). However, the series (\ref{ser2}) contains
a big "renormalization freedom" since we can change in arbitrary
way the higher orders corrections 
${\bf S}_{(2)}(X,T)$, ${\bf S}_{(3)}(X,T)$, 
$\dots$, ${\bf n}_{(1)}(X,T)$, ${\bf n}_{(2)}(X,T)$, $\dots$
and then adjust the functions $\bm{\tilde \Psi}_{(k)}$
in the appropriate way.

 Let us now "fix the normalization" of the solution (\ref{ser2})
in the following way:

\vspace{0.5cm}

 We require that the first term

$$\Phi^{i} \left( {{\bf S}(X,T,\epsilon) \over \epsilon}
\, + \, \bm{\theta}, \,\,\, {\bf S}_{X}(X,T,\epsilon), \,\,\,  
{\bf S}_{T}(X,T,\epsilon), \,\,\, {\bf n} (X,T,\epsilon) \right) $$
of (\ref{ser2}) gives "the best approximation" of the
asymptotic solution (\ref{ser2}) by the modulated $m$-phase 
solution of (\ref{insyst}). To be more precise, we require that the 
rest of the series (\ref{ser2}) is orthogonal (at every $X$ and $T$)
to the vectors 
$\bm{\Phi}_{\theta^{\alpha}}(\bm{\theta}, \, X, \, T)$, \,
$\bm{\Phi}_{n^{l}}(\bm{\theta}, \, X, \, T)$
"tangent" to $\Lambda$ at the "point"
$[{\bf S}(X,T,\epsilon), {\bf n}(X,T,\epsilon)]$.

\vspace{0.5cm}

 Thus we will put now the conditions

\begin{equation}
\label{fixnorm1}
\int_{0}^{2\pi} \!\! \dots \int_{0}^{2\pi} \sum_{i=1}^{n} \, 
\Phi^{i}_{\theta^{\alpha}} (\bm{\theta}, \, {\bf S}_{X}, \,
{\bf S}_{T}, \, {\bf n}) \,\,\, \left[ \sum_{k=1}^{\infty} \,
{\tilde \Psi}^{i}_{(k)} (\bm{\theta}, \, X, \, T) \,\,
\epsilon^{k} \right] \,\,\, {d^{m} \theta \over (2\pi)^{m}}
\,\,\, = \,\,\, 0
\end{equation}

\begin{equation}
\label{fixnorm2}
\int_{0}^{2\pi} \!\! \dots \int_{0}^{2\pi} \sum_{i=1}^{n} \,
\Phi^{i}_{n^{l}} (\bm{\theta}, \, {\bf S}_{X}, \,
{\bf S}_{T}, \, {\bf n}) \,\,\, \left[ \sum_{k=1}^{\infty} \,
{\tilde \Psi}^{i}_{(k)} (\bm{\theta}, \, X, \, T) \,\,
\epsilon^{k} \right] \,\,\, {d^{m} \theta \over (2\pi)^{m}}
\,\,\, = \,\,\, 0
\end{equation}
(for all $\epsilon$).

\vspace{0.5cm}

 As we saw above, the functions 
$\bm{\tilde \Psi}_{(k)} (\bm{\theta},X,T)$ are defined modulo 
the linear combinations of the functions
$\bm{\Phi}_{\theta^{\alpha}} (\bm{\theta}, \, {\bf S}_{(0)X}, \,
{\bf S}_{(0)T}, \, {\bf n}_{(0)} )$,
$\bm{\Phi}_{n^{l}} (\bm{\theta}, \, {\bf S}_{(0)X}, \,
{\bf S}_{(0)T}, \, {\bf n}_{(0)} )$ which are linearly 
independent for the full regular family of $m$-phase solutions
of (\ref{insyst}). Using this fact it's not difficult to prove
that the conditions (\ref{fixnorm1})-(\ref{fixnorm2}) fix
uniquely all the terms ${\bf S}_{(k)}$, ${\bf n}_{(k)}$,
and $\bm{\tilde \Psi}_{(k)} (\bm{\theta},X,T)$ ($k \geq 0$)
for a given solution (\ref{ser2}).

\vspace{0.5cm}

 Let us say now that the choice of the normalization
(\ref{fixnorm1})-(\ref{fixnorm2}) is not unique. In particular,
it depends on the choice of the variables $\varphi^{i}(x,t)$
for the "vector" ($i > 1$) systems (\ref{insyst}). We will
speak about the normalization (\ref{fixnorm1})-(\ref{fixnorm2})
as about one possible way to fix the functions
$\bm{\tilde \Psi}_{(k)} (\bm{\theta},X,T)$.

\vspace{0.5cm}

 The solution $\bm{\phi} \,\, (\bm{\theta},X,T)$ can now be 
defined as the asymptotic solution of the system (\ref{insyst1}) 
having the form (\ref{ser2}) which satisfies the conditions

$$\int_{0}^{2\pi} \!\!\! \dots \int_{0}^{2\pi} \sum_{i=1}^{n} \,
\Phi^{i}_{\theta^{\alpha}} (\bm{\theta}, \, {\bf S}_{X}, \,  
{\bf S}_{T}, \, {\bf n}) \,\,\,\,\,  
\phi^{i} \left(\bm{\theta} \, - \, {{\bf S} \over \epsilon}, \,
X, \, T, \, \epsilon \right)  \,\,\, 
{d^{m} \theta \over (2\pi)^{m}} \,\,\, = \,\,\, 0 $$

$$\int_{0}^{2\pi} \!\!\! \dots \int_{0}^{2\pi} \sum_{i=1}^{n} \,
\Phi^{i}_{n^{l}} (\bm{\theta}, {\bf S}_{X},  
{\bf S}_{T}, {\bf n}) \, \left[
\phi^{i} \left(\bm{\theta}  -  {{\bf S} \over \epsilon}, \,
X,  T,  \epsilon \right) \, - \,
\Phi^{i} (\bm{\theta}, {\bf S}_{X}, 
{\bf S}_{T}, {\bf n}) \right] \,
{d^{m} \theta \over (2\pi)^{m}} \, = \, 0 $$
(for all $\epsilon$).

\vspace{0.5cm}

 To find a solution in the form (\ref{ser2}) we substitute the
series (\ref{ser2}) in the system (\ref{insyst1}) and try to
find the functions ${\tilde \Psi}^{i}_{(k)} (\bm{\theta},X,T)$
from the linear systems

\begin{equation}
\label{novksyst}
{\hat L}^{i}_{j[{\bf S}_{(0)}, {\bf n}_{(0)}]} \,\,
{\tilde \Psi}^{j}_{(k)} (\bm{\theta},X,T) \,\,\, = \,\,\,
{\tilde f}^{i}_{(k)} (\bm{\theta},X,T)
\end{equation}
analogous to (\ref{ksyst}).

 The functions ${\tilde f}^{i}_{(k)} (\bm{\theta},X,T)$
are different from the functions 
$f^{i}_{(k)} (\bm{\theta},X,T)$ since we included "a part of
$\epsilon$-dependence" in the first term of (\ref{ser2}).
The functions ${\bf {\tilde f}}_{(k)} (\bm{\theta},X,T)$
should be orthogonal to the functions 
${\bf \kappa}^{(q)}_{[{\bf S}_{(0)}, {\bf n}_{(0)}]} 
(\bm{\theta},X,T)$, $q = 1, \dots, m+s$ and we use the conditions
(\ref{fixnorm1})-(\ref{fixnorm2}) for the recurrent determination
of ${\bf {\tilde \Psi}}_{(k)} (\bm{\theta},X,T)$. 
The functions $S^{\alpha}_{(k)}(X,T)$, $n^{l}_{(k)}(X,T)$ 
are used now to provide the resolvability of the systems 
(\ref{novksyst}) in all orders of $\epsilon$.

 Let us investigate now the systems arising on the functions
${\bf S}_{(0)}(X,T)$, ${\bf S}_{(1)}(X,T)$, $\dots$,
${\bf n}_{(0)}(X,T)$, ${\bf n}_{(1)}(X,T)$, $\dots$ .
Let us note that we have in our notations 
${\bf S}(X,T) \, = \, {\bf S}_{(0)}(X,T)$,
$\bm{\theta}_{0}(X,T) \, = \, {\bf S}_{(1)}(X,T)$. We saw
in Lemma 1.1 that the function $\bm{\theta}_{0}(X,T)$
does not appear in the solvability conditions of the system
(\ref{firstappr}). For the asymptotic solution written in
the form (\ref{ser2}) with the normalization conditions
(\ref{fixnorm1})-(\ref{fixnorm2}) we can prove here even 
stronger statement.

\vspace{0.5cm}

{\bf Lemma 2.1}

{\it The functions $S^{\alpha}_{(k)}(X,T)$, $n^{l}_{(k)}(X,T)$
do not appear in the expression for the discrepancy
${\bf {\tilde f}}_{(k)} (\bm{\theta},X,T)$ and do not affect
the solution ${\bf {\tilde \Psi}}_{(k)} (\bm{\theta},X,T)$. 
}

\vspace{0.5cm}

 Proof.

 The way how we get the discrepancy 
${\bf {\tilde f}}_{(k)} (\bm{\theta},X,T)$ can be described
as follows:

1) We substitute the solution (\ref{ser2}) in the system
(\ref{insyst1});

2) After "making all differentiations" we can omit the
argument shift ${\bf S}(X,T,\epsilon)/\epsilon$ in all functions
depending on $\bm{\theta}$;

3) Then collecting together all the terms of order 
$\epsilon^{k}$ we get the system ({\ref{novksyst}).

 It's not difficult to see that in this approach the $k$-th
order of $\epsilon$ containing the functions 
${\bf S}_{(k)}(X,T)$ or ${\bf n}_{(k)}(X,T)$ have the form

$${d F^{i} \over d k^{\alpha}} \,\, S^{\alpha}_{(k)X} 
\,\,\,\,\, , \,\,\,\,\, 
{d F^{i} \over d \omega^{\alpha}} \,\, S^{\alpha}_{(k)T}
\,\,\,\,\, , \,\,\,\,\, {\rm and} \,\,\,\,\, 
{d F^{i} \over d n^{l}} \,\, n^{l}_{(k)} $$
where $F^{i}$ are the constraints (\ref{phsyst}) defining
the $m$-phase solutions of (\ref{insyst}). The derivatives
$d /d k^{\alpha}$, $d / d \omega^{\alpha}$, $d / d n^{l}$
here are the total derivatives including the dependence of the
functions 
$\Phi^{i} (\bm{\theta}, {\bf k}, \bm{\omega}, {\bf n})$
on the corresponding parameters view the form of (\ref{ser2}).
However, all these derivatives are identically equal to zero
on the family $\Lambda$, so we get the first part of the Lemma.
To finish the proof we have to note also that the values
${\bf S}_{(k)}(X,T)$, ${\bf n}_{(k)}(X,T)$ do not appear
in the $k$-th order of $\epsilon$ of the normalization 
conditions (\ref{fixnorm1})-(\ref{fixnorm2}) either.

{\hfill Lemma 2.1 is proved.}

\vspace{0.5cm}

 We can formulate now the procedure in the following form:

\vspace{0.5cm}

 We try to solve the systems (\ref{novksyst}) recursively in
all orders and find the functions 
${\bf {\tilde \Psi}}_{(k)} (\bm{\theta},X,T)$ satisfying
the conditions (\ref{fixnorm1})-(\ref{fixnorm2}). At each
$k$-th step of our procedure we get a system on the functions
${\bf S}_{(k-1)}(X,T)$, ${\bf n}_{(k-1)}(X,T)$ from the
solvability conditions of (\ref{novksyst}). The full set of
Whitham solutions will then be parameterized by the set of all
functions $\{{\bf S}_{(k)}(X,T),{\bf n}_{(k)}(X,T)\}$,
$k \geq 0$ satisfying the conditions of solvability
of systems (\ref{novksyst}).

\vspace{0.5cm}

 Now let us investigate the systems arising on the functions
${\bf S}_{(k-1)}(X,T)$, ${\bf n}_{(k-1)}(X,T)$
from the solvability conditions of (\ref{novksyst})

\begin{equation}
\label{newortcond}
\int_{0}^{2\pi} \!\!\! \dots \int_{0}^{2\pi} \,\,
\kappa^{(q)}_{[{\bf S}_{0},{\bf n}_{0}]i} 
(\bm{\theta},X,T) \,\, {\tilde f}^{i}_{(k)}
(\bm{\theta},X,T) \,\,\, {d^{m} \theta \over (2\pi)^{m}} 
\,\,\,\,\, \equiv \,\,\,\,\, 0 \,\,\,\,\, ,
\,\,\,\,\,\,\,\, q = 1, \dots , m + s
\end{equation}
in the $k$-th order of $\epsilon$. We note first that the
solvability conditions of (\ref{novksyst}) for $k = 1$ give a 
nonlinear system on the functions 
${\bf S}_{(0)}(X,T)$, ${\bf n}_{(0)}(X,T)$.
It's not difficult to prove the following Lemma:

\vspace{0.5cm}

{\bf Lemma 2.2.}

{\it The system arising for $k = 1$ coincides with the Whitham 
system for the functions 
${\bf S}_{(0)}(X,T)$, ${\bf n}_{(0)}(X,T)$.
}

\vspace{0.5cm}

 Proof.

 Indeed, using Lemma 2.1 it's not difficult to see that the
discrepancy ${\bf f}_{(1)} (\bm{\theta},X,T)$ differs from
${\bf {\tilde f}}_{(1)} (\bm{\theta},X,T)$ just by the terms
containing derivatives $\bm{\theta}_{0X}$ and $\bm{\theta}_{0T}$.
However, according to Lemma 1.1 these terms do not affect
the orthogonality conditions (\ref{ortcond}) which then coincide
with orthogonality conditions for 
${\bf {\tilde f}}_{(1)} (\bm{\theta},X,T)$.

{\hfill Lemma 2.2 is proved.}

\vspace{0.5cm}

 The Whitham system (orthogonality conditions) for the functions
${\bf S}_{(0)}(X,T)$, ${\bf n}_{(0)}(X,T)$ can be written in the
following general form:

$$W^{(q)} \left[{\bf S}_{(0)}, {\bf n}_{(0)} \right] 
\,\,\,\,\,\,\,\, \equiv  \,\,\,\,\,\,\,\, A^{(q)}_{\alpha} 
\left({\bf S}_{(0)X}, {\bf S}_{(0)T}, {\bf n}_{(0)} \right) \,\,
S^{\alpha}_{(0)TT} \,\,\, + $$
$$+ \,\,\, B^{(q)}_{\alpha}
\left({\bf S}_{(0)X}, {\bf S}_{(0)T}, {\bf n}_{(0)} \right) \,\,
S^{\alpha}_{(0)XT} \,\,\, + \,\,\, C^{(q)}_{\alpha}
\left({\bf S}_{(0)X}, {\bf S}_{(0)T}, {\bf n}_{(0)} \right) \,\,
S^{\alpha}_{(0)XX} \,\,\, + $$
\begin{equation}
\label{wqsyst}
+ \,\,\, G^{(q)}_{l} 
\left({\bf S}_{(0)X}, {\bf S}_{(0)T}, {\bf n}_{(0)} \right) \,\,
n^{l}_{(0)T} \,\,\, + \,\,\, H^{(q)}_{l}
\left({\bf S}_{(0)X}, {\bf S}_{(0)T}, {\bf n}_{(0)} \right) \,\,
n^{l}_{(0)X} \,\,\, = \,\,\, 0 
\end{equation}
with some functions $A^{(q)}_{\alpha}$, $B^{(q)}_{\alpha}$,
$C^{(q)}_{\alpha}$, $G^{(q)}_{l}$, $H^{(q)}_{l}$, $\,\,$
$q \, = \, 1, \dots , m+s$, $\,\,$ $\alpha \, = \, 1, \dots , m$,
$\,\,$ $l \, = \, 1, \dots , s$ .

\vspace{0.5cm}

 Let us prove now the following Lemma about the systems on
${\bf S}_{(k)}$,  ${\bf n}_{(k)}$, $\,\,$ $k \geq 1$.

\vspace{0.5cm}

{\bf Lemma 2.3.}

{\it The orthogonality conditions of 
${\bf {\tilde f}}_{(k+1)} (\bm{\theta},X,T)$ to the functions
$\bm{\kappa}^{(q)}_{[{\bf S}_{(0)},{\bf n}_{(0)}]}$ give the
following linear systems for the functions
${\bf S}_{(k)}(X,T)$,  
${\bf n}_{(k)}(X,T)$, $\,\,$ $k \geq 1$:

$$\int \int {\delta W^{(q)}(X,T) \over \delta S^{\alpha}_{(0)}
(X^{\prime},T^{\prime}) } \,\,
S^{\alpha}_{(k)} (X^{\prime},T^{\prime}) \,\, d \, X^{\prime}
\,\, d \, T^{\prime} \,\,\, + \,\,\,
\int \int {\delta W^{(q)}(X,T) \over \delta n^{l}_{(0)}
(X^{\prime},T^{\prime}) } \,\, 
n^{l}_{(k)} (X^{\prime},T^{\prime}) \,\, d \, X^{\prime}
\,\, d \, T^{\prime} \,\,\, = $$
$$= \,\,\, V^{(q)}_{(k)} 
\left[ {\bf S}_{(0)}, \dots , {\bf S}_{(k-1)},
{\bf n}_{(0)}, \dots , {\bf n}_{(k-1)} \right] \, (X,T) $$
}

\vspace{0.5cm}

 In other words, the functions 
${\bf S}_{(k)}(X,T)$,  ${\bf n}_{(k)}(X,T)$
satisfy the linearized Whitham system on the functions
${\bf S}_{(0)}(X,T)$,  ${\bf n}_{(0)}(X,T)$
with additional right-hand part depending on the functions
${\bf S}_{(0)}$, $\dots$, ${\bf S}_{(k-1)}$,
${\bf n}_{(0)}$, $\dots$, ${\bf n}_{(k-1)}$.

\vspace{0.5cm}

 Proof.

 Let us look at the terms in 
${\bf {\tilde f}}_{(k+1)} (\bm{\theta},X,T)$ which contain
the functions ${\bf S}_{(k)}(X,T)$,  ${\bf n}_{(k)}(X,T)$ :

1) As we proved in Lemma 2.1 the functions 
${\tilde f}^{i}_{(1)} (\bm{\theta},X,T)$ contain only the terms
depending on ${\bf S}_{(0)}$,  ${\bf n}_{(0)}$. Easy to
see that the functions 
${\tilde f}^{i}_{(k+1)} (\bm{\theta},X,T)$ will then contain 
the terms

$$\int \int {\delta {\tilde f}^{i}_{(1)} (\bm{\theta},X,T) 
\over \delta S^{\alpha}_{(0)}
(X^{\prime},T^{\prime}) } \,\,
S^{\alpha}_{(k)} (X^{\prime},T^{\prime}) \,\, d \, X^{\prime}
\,\, d \, T^{\prime} \,\,\, + \,\,\,
\int \int {\delta {\tilde f}^{i}_{(1)} (\bm{\theta},X,T)
\over \delta n^{l}_{(0)}
(X^{\prime},T^{\prime}) } \,\,
n^{l}_{(k)} (X^{\prime},T^{\prime}) \,\, d \, X^{\prime}  
\,\, d \, T^{\prime} $$
according to the form of the first term in (\ref{ser2}).

2) There are terms containing the functions 
${\bf S}_{(k)}$,  ${\bf n}_{(k)}$ and the function
${\bf {\tilde \Psi}}_{(1)} (\bm{\theta},X,T)$. All such terms
can be written in the form:

$$- \! \int\!\!\int \left[
S^{\alpha}_{(k)} (X^{\prime},T^{\prime}) \,
{\delta {\hat L}^{i}_{[{\bf S}_{0},{\bf n}_{0}]j} (X,T)
\over \delta S^{\alpha}_{(0)}
(X^{\prime},T^{\prime}) } \, + \,
n^{l}_{(k)} (X^{\prime},T^{\prime}) \,
{\delta {\hat L}^{i}_{[{\bf S}_{0},{\bf n}_{0}]j} (X,T)
\over \delta n^{l}_{(0)} (X^{\prime},T^{\prime}) } \right] 
\! {\tilde \Psi}^{j}_{(1)} (\bm{\theta},X,T) \,\,
d X^{\prime} \, d T^{\prime} $$
where ${\hat L}^{i}_{j[{\bf S}_{0},{\bf n}_{0}]}$ is the linear
operator (\ref{linsyst}) given by the linearization of the
system (\ref{phsyst}) on the family of $m$-phase solutions.

3) There will the terms of the form

$$- \, \int \dots \int 
{\delta^{2} F^{i}_{[{\bf S}_{0},{\bf n}_{0}]} (\bm{\theta},X,T) 
\over \delta S^{\alpha}_{(0)} (X^{\prime},T^{\prime})  \,
\delta S^{\beta}_{(0)} (X^{\prime\prime},T^{\prime\prime})}
\,\,\, S^{\alpha}_{(k)} (X^{\prime},T^{\prime})  \,\,
S^{\beta}_{(1)} (X^{\prime\prime},T^{\prime\prime}) \,\,
d X^{\prime} \, d T^{\prime} \,
d X^{\prime\prime} \, d T^{\prime\prime} \,\, - \,\, \dots $$
where $F^{i}_{[{\bf S}_{0},{\bf n}_{0}]} (\bm{\theta},X,T)$
is the left-hand part of the system (\ref{phsyst}).

 However, the sum of all such terms is equal to zero since they 
all correspond to the expansion of the "shift of parameters"
${\bf S}_{(0)}(X,T)$,  ${\bf n}_{(0)}(X,T)$ on the family
$\Lambda$ where we have $F^{i}(\bm{\theta},X,T) \, \equiv \, 0$
identically.

\vspace{0.3cm}

 Let us look now at the terms in the orthogonality conditions

\begin{equation}
\label{ktf}
\int_{0}^{2\pi} \!\! \dots \int_{0}^{2\pi} \,\, 
\kappa^{(q)}_{[{\bf S}_{0},{\bf n}_{0}]i}
(\bm{\theta},X,T) \,\,\,\,\, 
{\tilde f}^{i}_{(k+1)} (\bm{\theta},X,T) \,\,\,\,\,
{d^{m} \theta \over (2\pi)^{m}} \,\,\,\,\, = \,\,\,\,\, 0 
\end{equation}
containing the terms 1)-2).

 We have identically

\begin{equation}
\label{kf}
\int_{0}^{2\pi} \!\! \dots \int_{0}^{2\pi} \,\,
\kappa^{(q)}_{[{\bf S}_{0},{\bf n}_{0}]i}
(\bm{\theta},X,T) \,\,\,\,\,
L^{i}_{[{\bf S}_{0},{\bf n}_{0}]j} 
(\bm{\theta},\bm{\theta}^{\prime},X,T)
\,\,\,\,\,
{d^{m} \theta \over (2\pi)^{m}} \,\,\,\,\, 
\equiv \,\,\,\,\, 0 
\end{equation}
on $\Lambda$ (where 
$L^{i}_{[{\bf S}_{0},{\bf n}_{0}]j}
(\bm{\theta},\bm{\theta}^{\prime},X,T)$
is the "core" of the operator 
${\hat L}^{i}_{[{\bf S}_{0},{\bf n}_{0}]j} (X,T)$).

 Easy to see then that the inner product of $\bm{\kappa}^{(q)}$
with the terms 2) is equal to

$$\int_{0}^{2\pi} \!\!\!\!\! \dots \! \int_{0}^{2\pi} 
\!\!\! \int \!\! \int \left[
S^{\alpha}_{(k)} (X^{\prime},T^{\prime})  \,
{\delta \kappa^{(q)}_{[{\bf S}_{0},{\bf n}_{0}]i}
(\bm{\theta},X,T) \over \delta S^{\alpha}_{(0)}
(X^{\prime},T^{\prime}) } \, + \,
n^{l}_{(k)} (X^{\prime},T^{\prime}) \,
{\delta \kappa^{(q)}_{[{\bf S}_{0},{\bf n}_{0}]i}
(\bm{\theta},X,T)
\over \delta n^{l}_{(0)} (X^{\prime},T^{\prime}) } \right]
d X^{\prime} d T^{\prime} \,\, \times $$
$$\times \,\,\, {\hat L}^{i}_{[{\bf S}_{0},{\bf n}_{0}]j} (X,T)
\,\,\, {\tilde \Psi}^{j}_{(1)} (\bm{\theta},X,T) \,\,\,
{d^{m} \theta \over (2\pi)^{m}} $$

 It's not difficult to see now that the terms of orthogonality
conditions containing the terms 1)-2) can be written together
in the form

$$\int \!\! \int S^{\alpha}_{(k)} (X^{\prime},T^{\prime})  
\,\, {\delta \over \delta S^{\alpha}_{(0)}
(X^{\prime},T^{\prime}) } \,\, \langle
\bm{\kappa}^{(q)}_{[{\bf S}_{0},{\bf n}_{0}]} \,\, \cdot \,\,
{\bf {\tilde f}}_{(1)}[{\bf S}_{0},{\bf n}_{0}] \rangle
(X,T) \,\,\, d X^{\prime} \, d T^{\prime} \,\,\, + $$
$$+ \,\,\, \int \!\! \int 
n^{l}_{(k)} (X^{\prime},T^{\prime})
\,\, {\delta \over \delta n^{l}_{(0)}
(X^{\prime},T^{\prime}) } \,\, \langle
\bm{\kappa}^{(q)}_{[{\bf S}_{0},{\bf n}_{0}]} \,\, \cdot \,\,
{\bf {\tilde f}}_{(1)}[{\bf S}_{0},{\bf n}_{0}] \rangle
(X,T) \,\,\, d X^{\prime} \, d T^{\prime} $$   
(where $< \dots \cdot \dots >$ is the inner product of
$\bm{\kappa}^{(q)}$ and ${\bf {\tilde f}}_{(1)}$).

 All the other terms in the orthogonality conditions
(\ref{ktf}) are the smooth functionals of
${\bf S}_{(0)}$, $\dots$, ${\bf S}_{(k-1)}$,
${\bf n}_{(0)}$, $\dots$, ${\bf n}_{(k-1)}$,
so we get the statement of the Lemma.

{\hfill Lemma 2.3 is proved.}

\vspace{0.5cm}

 Let us consider now the systems (\ref{insyst}) satisfying
the special non-degeneracy conditions. Namely, we will assume
that the corresponding Whitham system (\ref{wqsyst}) 
can be resolved w.r.t. to the highest $T$-derivatives of
the functions ${\bf S}_{(0)}(X,T)$, ${\bf n}_{(0)}(X,T)$
and written in the "evolution" form:

$$S^{\alpha}_{(0)TT} \,\,\, = \,\,\, M^{\alpha}_{(0)\beta}
\left({\bf S}_{(0)X}, {\bf S}_{(0)T}, {\bf n}_{(0)} \right) \,\,
S^{\beta}_{(0)XX} \,\,\, + \,\,\, N^{\alpha}_{(0)\beta}
\left({\bf S}_{(0)X}, {\bf S}_{(0)T}, {\bf n}_{(0)} \right) \,\,
S^{\beta}_{(0)TX} \,\,\, +$$
\begin{equation}
\label{snsyst1}
+ \,\,\, P^{\alpha}_{(0)p}
\left({\bf S}_{(0)X}, {\bf S}_{(0)T}, {\bf n}_{(0)} \right) \,\,
n^{p}_{(0)X} \,\,\,\,\, , \,\,\,\,\,\,\,\, 
\alpha \, = \, 1, \dots , m 
\end{equation}

$$n^{l}_{(0)T} \,\,\, = \,\,\, T^{l}_{(0)\beta}
\left({\bf S}_{(0)X}, {\bf S}_{(0)T}, {\bf n}_{(0)} \right) \,\,
S^{\beta}_{(0)XX} \,\,\, + \,\,\, L^{l}_{(0)\beta}
\left({\bf S}_{(0)X}, {\bf S}_{(0)T}, {\bf n}_{(0)} \right) \,\,
S^{\beta}_{(0)TX} \,\,\, +$$
\begin{equation}
\label{snsyst2}
+ \,\,\, R^{l}_{(0)p}    
\left({\bf S}_{(0)X}, {\bf S}_{(0)T}, {\bf n}_{(0)} \right) \,\,
n^{p}_{(0)X} \,\,\,\,\, , \,\,\,\,\,\,\,\,
l \, = \, 1, \dots , s
\end{equation}

 After the introduction of the variables 
$k^{\alpha}_{(0)} = S^{\alpha}_{(0)X}$ , 
$\omega^{\alpha}_{(0)} = S^{\alpha}_{(0)T}$
we can write the system (\ref{snsyst1})-(\ref{snsyst2})
in the form

$$k^{\alpha}_{(0)T} \,\,\, = \,\,\, \omega^{\alpha}_{(0)X} $$

$$\omega^{\alpha}_{(0)T} \,\,\, = \,\,\, M^{\alpha}_{(0)\beta}
\left({\bf k}_{(0)}, \bm{\omega}_{(0)}, {\bf n}_{(0)} \right) \,\,
k^{\beta}_{(0)X} \,\,\, + \,\,\, N^{\alpha}_{(0)\beta}
\left({\bf k}_{(0)}, \bm{\omega}_{(0)}, {\bf n}_{(0)} \right) \,\,
\omega^{\beta}_{(0)X} \,\,\, +$$
\begin{equation}
\label{kon}
+ \,\,\, P^{\alpha}_{(0)p}
\left({\bf k}_{(0)}, \bm{\omega}_{(0)}, {\bf n}_{(0)} \right) \,\,
n^{p}_{(0)X} 
\end{equation}

$$n^{l}_{(0)T} \,\,\, = \,\,\, T^{l}_{(0)\beta}
\left({\bf k}_{(0)}, \bm{\omega}_{(0)}, {\bf n}_{(0)} \right) \,\,
k^{\beta}_{(0)X} \,\,\, + \,\,\, L^{l}_{(0)\beta}
\left({\bf k}_{(0)}, \bm{\omega}_{(0)}, {\bf n}_{(0)} \right) \,\,
\omega^{\beta}_{(0)X} \,\,\, +$$
$$ + \,\,\, R^{l}_{(0)p}
\left({\bf k}_{(0)}, \bm{\omega}_{(0)}, {\bf n}_{(0)} \right) \,\,
n^{p}_{(0)X} $$
i.e. in the form (\ref{HTsyst}).

\vspace{0.5cm}

 We will consider now the hyperbolic systems (\ref{kon}), i.e.
such that the matrix

$$V^{\nu}_{\mu} ({\bf k}_{(0)}, \bm{\omega}_{(0)}, {\bf n}_{(0)})
 \,\,\, = \,\,\,
\left( \begin{array}{c|c|c}
0 & I_{m} & 0 \cr
\hline
M_{(0)} & N_{(0)} & P_{(0)} \cr
\hline
T_{(0)} & L_{(0)} & R_{(0)}
\end{array} \right) $$ 
has exactly $N = 2m+s$ real eigen-values
with $N$ linearly independent real eigen-vectors. For
hyperbolic systems (\ref{kon}) it's natural to consider the
Cauchy problem with the smooth initial data
${\bf k}_{(0)}(X,0)$, $\bm{\omega}_{(0)}(X,0)$,
${\bf n}_{(0)}(X,0)$ (or ${\bf S}_{(0)}(X,0)$,
${\bf S}_{(0)T}(X,0)$, ${\bf n}_{(0)}(X,0)$). The smooth
solution of (\ref{kon}) exists in general up to some finite
time $T_{0}$ until the breakdown occurs. So we can write the 
zero (global in $X$) approximation for the solution
(\ref{ser2}) just in the time interval where we have a smooth
solution of the Whitham system. Using Lemma 2.3 it's not 
difficult to prove then the following Lemma:

\vspace{0.5cm}

{\bf Lemma 2.4}

{\it For non-degenerate hyperbolic Whitham system (\ref{kon})
and the global solution ${\bf S}_{(0)}(X,T)$, 
${\bf n}_{(0)}(X,T)$ defined on the interval $[0, T_{0}]$
the higher orders approximations in (\ref{ser2}) are all defined
for all $X$ and $T \in [0, T_{0})$ and are parameterized by
the initial values
${\bf S}_{(k)}(X,0)$,
${\bf S}_{(k)T}(X,0)$, ${\bf n}_{(k)}(X,0)$.
\footnote{Let us remind that we assume that all systems 
(\ref{novksyst}) are solvable if the corresponding orthogonality
conditions are satisfied.}
}

\vspace{0.5cm}

 Proof.

 Indeed, as follows from Lemma 2.3 the functions 
${\bf S}_{(k)}(X,T)$, ${\bf n}_{(k)}(X,T)$
are defined by the initial values
${\bf S}_{(k)}(X,0)$,
${\bf S}_{(k)T}(X,0)$, ${\bf n}_{(k)}(X,0)$
and can be found from the linear system using 
the characteristic directions of (\ref{kon}) 
(defined by ${\bf S}_{(0)}(X,T)$, ${\bf n}_{(0)}(X,T)$) 
provided that all the smooth solutions
${\bf S}_{(0)}(X,T)$,  $\dots$  ,  
${\bf S}_{(k-1)}(X,T)$,  ${\bf n}_{(0)}(X,T)$ 
 $\dots$  ,   ${\bf n}_{(k-1)}(X,T)$ and
$\bm{{\tilde \Psi}}_{(1)}(\bm{\theta},X,T)$,  $\dots$ ,  
 $\bm{{\tilde \Psi}}_{(k-2)}(\bm{\theta},X,T)$
exist on the interval $[0, T_{0})$. According to Lemma 2.1
and Lemma 2.3 we can find then the functions
${\tilde \Psi}^{i}_{(k-1)}(\bm{\theta},X,T)$ 
which are the local expressions (in $X$ and $T$) of
${\bf S}_{(0)}(X,T)$,  $\dots$  , 
${\bf S}_{(k-1)}(X,T)$,  ${\bf n}_{(0)}(X,T)$
 $\dots$  ,   ${\bf n}_{(k-1)}(X,T)$ and
their derivatives. Using the induction we then finish the
proof of the Lemma.

{\hfill Lemma 2.4 is proved.}

\vspace{0.5cm}

 According to Lemma 2.4 we can formulate now the following
statement:

\vspace{0.5cm}

 For the initial system (\ref{insyst}) having the non-degenerate
hyperbolic Whitham system (\ref{kon}) the corresponding Whitham
solutions (\ref{ser2}) (or (\ref{sol1})-(\ref{ser1})) are defined
by the initial values 
${\bf S}(X,0,\epsilon)$, ${\bf S}_{T}(X,0,\epsilon)$, 
${\bf n}(X,0,\epsilon)$ and exist in the time interval
$[0, T_{0})$ defined by the Whitham system (\ref{kon}) and
the initial data
${\bf S}_{(0)} (X,0) \, = \, {\bf S}(X,0,0)$,
${\bf S}_{(0)T} (X,0) \, = \, {\bf S}_{T}(X,0,0)$, and
${\bf n}_{(0)} (X,0) \, = \, {\bf n}(X,0,0)$.

\vspace{0.5cm}

\centerline{\bf The deformation procedure.}

\vspace{0.5cm}

 Let us note now that the series (\ref{ser2}) 
(or (\ref{sol1})-(\ref{ser1})) give in fact the one-parametric
formal solutions of (\ref{insyst}) with a parameter $\epsilon$.
Let us rewrite now the solutions (\ref{ser2}) in the form
which gives the concrete (formal) solution of (\ref{insyst})
and is not connected with the additional one-parametric
$\epsilon$-family including this given solution. We omit
now the $\epsilon$-dependence of functions 
${\bf S}(X,T,\epsilon)$, ${\bf n}(X,T,\epsilon)$
(or put formally $\epsilon = 1$) and say that the Whitham 
solution is defined now by functions
${\bf S}(X,T)$, ${\bf n}(X,T)$
determined by the initial values
${\bf S}(X,0)$, ${\bf S}_{T}(X,0)$, and ${\bf n}(X,0)$.
(Let us keep here the notations $X$ and $T$ for the spatial
and time coordinates just to emphasize that we consider the
"slow" functions ${\bf S}_{X}(X,T)$, ${\bf S}_{T}(X,T)$,
${\bf n}(X,T)$.)

 Thus, we define now the Whitham solution as the solution
of the system

\begin{equation}
\label{insyst2}
F^{i} (\bm{\varphi}, \, \bm{\varphi}_{T}, \, 
\bm{\varphi}_{X}, \, \bm{\varphi}_{TT}, \, 
\bm{\varphi}_{XT}, \, \bm{\varphi}_{XX}, \dots)
\, = \, 0 \,\,\,\,\,\,\,\, , \,\,\,\,\, i = 1, \dots, n
\end{equation}
having the form

$$\phi^{i} (\bm{\theta},X,T) \,\,\, = \,\,\,
\Phi^{i} \left( {\bf S}(X,T) \, + \, \bm{\theta}, \,\,\, 
{\bf S}_{X}(X,T), \,\,\, {\bf S}_{T}(X,T), \,\,\, 
{\bf n} (X,T) \right) \,\,\, + $$
\begin{equation}
\label{ser3}
+ \,\,\, \sum_{k \geq 1} \,
\Phi^{i}_{(k)} \left( {\bf S}(X,T) \, + \, \bm{\theta}, \,
X, \, T \right) 
\end{equation}
where the functions $\Phi^{i}_{(k)}$ :

\vspace{0.5cm}

1) Are $2\pi$-periodic with respect to each $\theta^{\alpha}$;

\vspace{0.3cm}

2) Have degree $k$ (introduced below);

\vspace{0.3cm}

3) Satisfy the normalization conditions
(\ref{fixnorm1})-(\ref{fixnorm2}) which will be written now
in the form:

\begin{equation}
\label{kort1}
\int_{0}^{2\pi} \!\!\! \dots \int_{0}^{2\pi} 
\sum_{i=1}^{n} \, \Phi^{i}_{\theta^{\alpha}}
\left( \bm{\theta}, \,\, {\bf S}_{X}, \, {\bf S}_{T}, \,
{\bf n} \right) \,\, \Phi^{i}_{(k)}
(\bm{\theta}, X, T) \,\,\, {d^{m} \theta \over (2\pi)^{m}} 
\,\,\,\,\, = \,\,\,\,\, 0 
\end{equation}

\begin{equation}
\label{kort2}
\int_{0}^{2\pi} \!\!\! \dots \int_{0}^{2\pi}     
\sum_{i=1}^{n} \, \Phi^{i}_{n^{l}}
\left( \bm{\theta}, \,\,
{\bf S}_{X}, \, {\bf S}_{T}, \,
{\bf n} \right) \,\,\, \Phi^{i}_{(k)}
(\bm{\theta}, X, T) \,\,\, {d^{m} \theta \over (2\pi)^{m}}
\,\,\,\,\, = \,\,\,\,\, 0 
\end{equation}
$k \geq 1$, $\,\,$ ($\alpha \, = \, 1, \dots, m$, $\,\,$ 
$l \, = \, 1, \dots, s$).

\vspace{0.5cm}

 Let us introduce now the gradation used for the formal 
expansion (\ref{ser3}). Namely, for the systems (\ref{insyst})
having the non-degenerate hyperbolic Whitham system 
(\ref{kon}) we put the following gradation on the functions
${\bf S}_{X}(X,T)$, ${\bf S}_{T}(X,T)$, ${\bf n}(X,T)$ and
their derivatives :

\vspace{0.5cm}

1) The functions $k^{\alpha}(X,T) \, = \, S^{\alpha}_{X}(X,T)$,
$\omega^{\alpha}(X,T) \, = \, S^{\alpha}_{T}(X,T)$, and
$n^{l}(X,T)$ have degree 0;

\vspace{0.3cm}

2) Every differentiation with respect to $X$ adds 1 to the degree
of the function;

\vspace{0.3cm}

3) The degree of the product of two functions having certain 
degrees is equal to the sum of their degrees.

\vspace{0.5cm}

 In other words, for the parameters 
${\bf U} = ({\bf k}, \bm{\omega}, {\bf n})$ we have the gradation 
rule of Dubrovin - Zhang type, i.e.

All the functions $f({\bf U})$ have degree $0$;

The derivatives $U^{\nu}_{kX}$ have degree $k$;

The degree of the product of functions having certain degrees    
is equal to the sum of their degrees.

\vspace{0.5cm}

 We put now the evolution conditions to the functions
${\bf S}(X,T)$, ${\bf n}(X,T)$ having the form :

\begin{equation}
\label{deform1}
S^{\alpha}_{TT} \,\,\, = \,\,\, \sum_{k\geq1} \,\,
\sigma^{\alpha}_{(k)} \, ({\bf k}, \, \bm{\omega}, \, 
{\bf n}, \, {\bf k}_{X}, \, \bm{\omega}_{X}, \, 
{\bf n}_{X}, \, \dots \, )
\end{equation}

\begin{equation}
\label{deform2}
n^{l}_{T} \,\,\, = \,\,\, \sum_{k\geq1} \,\,
\eta^{l}_{(k)} \, ({\bf k}, \, \bm{\omega}, \, {\bf n}, \,  
{\bf k}_{X}, \, \bm{\omega}_{X}, \, {\bf n}_{X}, \, \dots \, )
\end{equation}
where $\sigma^{\alpha}_{(k)}$, $\eta^{l}_{(k)}$ are general 
polynomials in derivatives ${\bf k}_{X}$, $\bm{\omega}_{X}$,
${\bf n}_{X}$, ${\bf k}_{XX}$, $\bm{\omega}_{XX}$,
${\bf n}_{XX}$, $\dots$ (with coefficients depending on
$({\bf k},\bm{\omega},{\bf n})$) having degree $k$.

\vspace{0.5cm}

 We now substitute the expansion (\ref{ser3}) in the system
(\ref{insyst2}) and use the relations 
(\ref{deform1})-(\ref{deform2}) to remove all the time
derivatives of parameters $({\bf k},\bm{\omega},{\bf n})$.
After that we can divide the system (\ref{insyst2}) into the 
terms of certain degrees and try to find recursively all the
terms $\bm{\Phi}_{(k)}(\bm{\theta},X,T)$ for 
$k = 1, 2, \dots$ .

\vspace{0.5cm}

 It is easy to see again that for any 
$\bm{\Phi}_{(k)}(\bm{\theta},X,T)$ we will have the linear 
system analogous to (\ref{ksyst}), (\ref{novksyst}), i.e.

\begin{equation}
\label{fksyst}
{\hat L}^{i}_{[{\bf S}(X,T),{\bf n}(X,T)]j} \,\,
\Phi^{i}_{(k)}(\bm{\theta},X,T) \,\,\, = \,\,\,
{\hat f}^{i}_{(k)} (\bm{\theta},X,T)
\end{equation}
where ${\bf {\hat f}}_{(k)} (\bm{\theta},X,T)$ is the
discrepancy having degree $k$ according to the definition 
above.

\vspace{0.5cm}

 We have to put again the orthogonality conditions

\begin{equation}
\label{fortcond}
\int_{0}^{2\pi} \!\!\! \dots \int_{0}^{2\pi} \,\,
\kappa^{(q)}_{[{\bf S},{\bf n}]i} (\bm{\theta},X,T)
\,\, {\hat f}^{i}_{(k)} (\bm{\theta},X,T)
\,\,\, {d^{m} \theta \over (2\pi)^{m}} \,\,\,\,\, 
\equiv \,\,\,\,\, 0
\end{equation}
on the functions ${\hat f}^{i}_{(k)} (\bm{\theta},X,T)$
and then find the unique $\bm{\Phi}_{(k)}(\bm{\theta},X,T)$
satisfying the normalization conditions 
(\ref{kort1})-(\ref{kort2}).

\vspace{0.8cm}

 It's not difficult to prove the following Lemma:

\vspace{0.8cm}

{\bf Lemma 2.5.}

{\it 
 1) For any system (\ref{insyst}) having the non-degenerate
hyperbolic Whitham system (\ref{kon}) the functions
$\sigma^{\alpha}_{(k)}$, $\eta^{l}_{(k)}$ are uniquely 
determined by the orthogonality conditions (\ref{fortcond})
in the order $k$.

 2) The functions $\sigma^{\alpha}_{(1)}$, $\eta^{l}_{(1)}$
give the Whitham system (\ref{snsyst1})-(\ref{snsyst2})
for the functions ${\bf S}_{(0)}(X,T)$, ${\bf n}_{(0)}(X,T)$.
\footnote{In fact, the functions ${\bf S}(X,T,\epsilon)$,
${\bf n}(X,T,\epsilon)$ introduced previously satisfy the 
full system (\ref{deform1})-(\ref{deform2}).}
}

\vspace{0.5cm}

 Proof.

 Indeed, using Lemma 2.1 it is easy to see that the functions
${\hat f}^{i}_{(1)}(\bm{\theta},X,T)$ coincide with the 
functions ${\tilde f}^{i}_{(1)}(\bm{\theta},X,T)$ introduced
in (\ref{novksyst}) after the replacement of functions
${\bf S}_{(0)}(X,T)$, ${\bf n}_{(0)}(X,T)$ by
${\bf S}(X,T)$, ${\bf n}(X,T)$. Comparing then the orthogonality
conditions (\ref{fortcond}) with (\ref{newortcond}) we get
the second part of the Lemma.

 To prove the first part we just note that all
$\sigma^{\alpha}_{(k)}$, $\eta^{l}_{(k)}$ arise in the
$k$-th order of system (\ref{insyst2}) "in the same way".
We can conclude then that the orthogonality conditions
(\ref{fortcond}) in the $k$-th order always contain the
functions $\sigma^{\alpha}_{(k)}$, $\eta^{l}_{(k)}$ 
in one particular way which coincides with the appearance
of $\sigma^{\alpha}_{(1)}$, $\eta^{l}_{(1)}$ in the
Whitham system arising for $k = 1$. From the definition
of the non-degenerate Whitham system we now obtain the first
part of the Lemma.

{\hfill Lemma 2.5 is proved.}

\vspace{0.5cm}

{\bf Definition 2.1.}

{\it We call the system (\ref{deform1})-(\ref{deform2})
or the equivalent system

$$k^{\alpha}_{T} \,\,\,\,\, = \,\,\,\,\, \omega^{\alpha}_{X} $$

\begin{equation}
\label{defwhsyst}
\omega^{\alpha}_{T} \,\,\, = \,\,\, \sum_{k\geq1} \,\,
\sigma^{\alpha}_{(k)} \, ({\bf k}, \, \bm{\omega}, \,
{\bf n}, \, {\bf k}_{X}, \, \bm{\omega}_{X}, \,
{\bf n}_{X}, \, \dots \, )
\end{equation}

$$n^{l}_{T} \,\,\, = \,\,\, \sum_{k\geq1} \,\,
\eta^{l}_{(k)} \, ({\bf k}, \, \bm{\omega}, \, {\bf n}, \,
{\bf k}_{X}, \, \bm{\omega}_{X}, \, {\bf n}_{X}, \, 
\dots \, ) $$
the deformation of the Whitham system 
(\ref{snsyst1})-(\ref{snsyst2}) (or (\ref{kon})).
}

\vspace{0.5cm}

 The functions $k^{\alpha}(X,T)$, $\omega^{\alpha}(X,T)$,
$n^{l}(X,T)$ are the "slow" functions of the variables
$x$ and $t$ and the system (\ref{defwhsyst}) gives the analog 
of the "low-dispersion" expansion in our case. The asymptotic
solutions (\ref{ser3}) of the initial system (\ref{insyst2})
are parameterized by the asymptotic solutions
$k^{\alpha}(X,T)$, $\omega^{\alpha}(X,T)$,
$n^{l}(X,T)$ of the system (\ref{defwhsyst}) and arbitrary
(constant) initial phases $\theta^{\alpha}_{0}$. As follows
from our considerations above the solutions (\ref{ser3})
give all the "particular solutions" (\ref{sol1})-(\ref{ser1}),
however, they do not contain the additional information
about the one-parametric $\epsilon$-family given by
(\ref{sol1})-(\ref{ser1}).

\vspace{0.8cm}

 Remark 1.

 Let us note that the full set of parameters of $m$-phase
solutions of (\ref{insyst}) is given by 
${\bf k}$, $\bm{\omega}$, ${\bf n}$, and $\bm{\theta}_{0}$.
However the functions $\bm{\theta}_{0}(X,T)$ do not present
as the parameters of solutions (\ref{ser3}) in this approach.
This shows in fact that the introduction of the functions
$\theta^{\alpha}_{0}(X,T)$ do not give "new" formal solutions
of (\ref{insyst}) and is responsible for the additional
$\epsilon$-dependence of one-parametric families
(\ref{sol1})-(\ref{ser1}). Here they are "absorbed" by the
total phase ${\bf S}(X,T)$ connected with the "particular"
formal solution of (\ref{insyst}).

\vspace{0.5cm}

 Remark 2.

In our consideration we fixed some functions
$\Phi^{i}(\bm{\theta}, \, {\bf k}, \, \bm{\omega}, \,
{\bf n})$ on the family $\Lambda$ (for each
$({\bf k}, \, \bm{\omega}, \, {\bf n})$) as having zero
initial phases. However, the choice of the functions
$\Phi^{i}(\bm{\theta}, \, {\bf k}, \, \bm{\omega}, \,
{\bf n})$ is not unique. In particular, the natural change 
of the functions \linebreak
$\Phi^{i}(\bm{\theta}, \, {\bf k}, \, \bm{\omega}, \,
{\bf n})$ on $\Lambda$ can be written in the form

\begin{equation}
\label{trans}
\bm{\Phi}^{\prime}(\bm{\theta}, \, {\bf k}, \, \bm{\omega}, \,
{\bf n}) \,\,\, = \,\,\,
\bm{\Phi} (\bm{\theta} \, + \, \bm{\theta}_{0}
({\bf k}, \, \bm{\omega}, \, {\bf n}) , \, {\bf k}, \, 
\bm{\omega}, \, {\bf n})
\end{equation}
where 
$\theta^{\alpha}_{0}({\bf k}, \, \bm{\omega}, \, {\bf n})$
are arbitrary smooth functions.

 It's not difficult to see also that the system 
(\ref{defwhsyst}) depends on the choice of the functions
$\Phi^{i}(\bm{\theta}, \, {\bf k}, \, \bm{\omega}, \,
{\bf n})$ in the high ($k \geq 2$) orders.

 Let us give here the definition given by B.A. Dubrovin
and Y. Zhang (\cite{DubrZhang1,DubrZhang2}) and connected
with the "equivalence" of different infinite (or finite) 
systems.

\vspace{0.5cm}

{\bf Definition 2.2} (B.A. Dubrovin, Y. Zhang).

{\it Consider the system of the form

\begin{equation}
\label{genusyst}
U^{\nu}_{T} \,\,\, = \,\,\, \sum_{k \geq 1} \, 
V^{\nu}_{(k)} \, ({\bf U}, \, {\bf U}_{X}, \,
{\bf U}_{XX}, \, \dots \, ) \,\,\,\,\, ,
\,\,\,\,\,\,\,\, \nu = 1, \dots, N
\end{equation} 
for arbitrary parameters $U^{\nu}$ where all
$V^{\nu}_{(k)}$ are smooth functions polynomial in
${\bf U}_{X}$, ${\bf U}_{XX}$, $\dots$ and having degree
$k$. We say that two systems (\ref{genusyst}) are connected
by the triviality transformation (or equivalent) if they are
connected by the formal substitution

$${\tilde U}^{\nu} \,\,\, = \,\,\, \sum_{k \geq 0} \,
{\tilde U}^{\nu}_{(k)} \, ({\bf U}, \, {\bf U}_{X}, \,
{\bf U}_{XX}, \, \dots \, ) $$
where all ${\tilde U}^{\nu}_{(k)}$ are smooth functions
polynomial in ${\bf U}_{X}$, ${\bf U}_{XX}$, $\dots$ 
and having degree $k$.
}

\vspace{0.5cm}

 Let us say actually that the definition of B.A. Dubrovin
and Y. Zhang is applied usually to the whole integrable
hierarchies and plays the important role in the classification
of integrable hierarchies having the form (\ref{genusyst}).
We will prove here the following Lemma:

\vspace{0.5cm}

{\bf Lemma 2.6.}

{\it The deformations of the Whitham system (\ref{defwhsyst})
written for the initial functions 
$\bm{\Phi}(\bm{\theta}, \, {\bf k}, \, \bm{\omega}, \,
{\bf n})$ and
$\bm{\Phi}^{\prime}(\bm{\theta}, \, {\bf k}, \, \bm{\omega}, \,
{\bf n})$ connected by the transformation (\ref{trans})
are equivalent in Dubrovin - Zhang sense.
}

\vspace{0.5cm}

 Proof.

 Let us prove first the following statement:

 For any transformation (\ref{trans}) there exists a unique
change of functions $S^{\alpha}(X)$, $n^{l}(X)$:

$$S^{\prime\alpha} \,\,\, = \,\,\, S^{\alpha} \,\, - \,\,
\theta^{\alpha}_{0} ({\bf k}, \, \bm{\omega}, \, {\bf n})
\,\, +  \,\, \sum_{k \geq 1} \, S^{\alpha}_{(k)} \,
({\bf k}, \, \bm{\omega}, \, {\bf n}, \, {\bf k}_{X}, 
\dots \,)$$
\begin{equation}
\label{sntrans}
n^{\prime l} \,\,\, = \,\,\, n^{l} \,\, +  \,\, 
\sum_{k \geq 1} \, n^{l}_{(k)} \, ({\bf k}, \, 
\bm{\omega}, \, {\bf n}, \, {\bf k}_{X}, \dots \,)
\end{equation}
such that:

\vspace{0.3cm}

 1) All ${\bf S}_{(k)}$, ${\bf n}_{(k)}$ are polynomial in
derivatives of $({\bf k}, \, \bm{\omega}, \, {\bf n})$
and have degree $k$;

\vspace{0.3cm}

 2) For any solution (\ref{ser3}) 
($\bm{\phi}_{[{\bf S},{\bf n}]} (\bm{\theta},X,T)$)
of (\ref{insyst2}) the functions \linebreak
$\Phi^{\prime i}({\bf S}^{\prime}(X,T) + \bm{\theta},
{\bf S}^{\prime}_{X}, {\bf S}^{\prime}_{T}, 
{\bf n}^{\prime})$
satisfy the normalization conditions

\begin{equation}
\label{snnorm1}
\int_{0}^{2\pi} \!\!\!\!\! \dots \! \int_{0}^{2\pi} 
\sum_{i = 1}^{n} \, \Phi^{\prime i}_{\theta^{\alpha}}
({\bf S}^{\prime} + \bm{\theta}, {\bf S}^{\prime}_{X}, 
{\bf S}^{\prime}_{T}, {\bf n}^{\prime}) \,\,
\phi^{i}_{[{\bf S},{\bf n}]} (\bm{\theta},X,T) \,\,\,
{d^{m} \theta \over (2\pi)^{m}} \,\,\,\,\, \equiv
\,\,\,\,\, 0
\end{equation}

\begin{equation}
\label{snnorm2}
\int_{0}^{2\pi} \!\!\!\!\! \dots \! \int_{0}^{2\pi} 
\sum_{i = 1}^{n} \Phi^{\prime i}_{n^{\prime l}}
({\bf S}^{\prime} + \bm{\theta}, {\bf S}^{\prime}_{X}, 
{\bf S}^{\prime}_{T}, {\bf n}^{\prime}) \left[
\phi^{i}_{[{\bf S},{\bf n}]} (\bm{\theta},X,T)  - 
\Phi^{\prime i} ({\bf S}^{\prime} + \bm{\theta},
{\bf S}^{\prime}_{X}, {\bf S}^{\prime}_{T}, 
{\bf n}^{\prime}) \right]
{d^{m} \theta \over (2\pi)^{m}} \, \equiv
\, 0
\end{equation}

 For the proof of this statement let us note first that we can 
always express all the time derivatives of ${\bf k}$, 
$\bm{\omega}$, and ${\bf n}$ using the system (\ref{defwhsyst})
in terms of $X$-derivatives of these functions. Using this
procedure we try to find the transformation (\ref{sntrans})
recursively in all degrees by substitution of (\ref{sntrans})
in (\ref{snnorm1})-(\ref{snnorm2}). It's not difficult to check
then that the functions $S^{\alpha}_{(k)}$, $n^{l}_{(k)}$
are defined in the $k$-th order of (\ref{snnorm1})-(\ref{snnorm2})
from a linear system. The matrix of this linear system coincides
with the Gram matrix of functions $\Phi_{\theta^{\alpha}}$,
$\Phi_{n^{l}}$ at every $X$ and $T$. Thus, for the full 
non-degenerate family of $m$-phase solutions of (\ref{insyst})
this system has a unique solution at every degree $k$. The
transformation (\ref{sntrans}) is evidently invertible in sense
of the infinite series (polynomial w.r.t. derivatives of
$({\bf k}, \, \bm{\omega}, \, {\bf n})$) so we can also express
the functions $({\bf S}, {\bf n})$ in terms of 
$({\bf S}^{\prime}, {\bf n}^{\prime})$.

 We can try to use now the functions 
$\Phi^{\prime i} ({\bf S}^{\prime} + \bm{\theta},
{\bf S}^{\prime}_{X}, {\bf S}^{\prime}_{T}, {\bf n}^{\prime})$ 
as the zero approximation in the 
$({\bf S}^{\prime}, {\bf n}^{\prime})$-expansion of the
corresponding solution (\ref{ser3}). It is not difficult
to see that the difference

$$\Phi^{\prime i} ({\bf S}^{\prime} + \bm{\theta},
{\bf S}^{\prime}_{X}, {\bf S}^{\prime}_{T}, {\bf n}^{\prime}) 
\,\,\, - \,\,\, 
\phi^{i}_{[{\bf S},{\bf n}]} (\bm{\theta},X,T)$$
can be represented as the infinite series polynomial w.r.t. 
derivatives of $({\bf k}, \, \bm{\omega}, \, {\bf n})$
and starting with the terms of degree 1. After the expression
of the functions $({\bf k}, \, \bm{\omega}, \, {\bf n})$
in terms of 
$({\bf k}^{\prime}, \, \bm{\omega}^{\prime}, \, 
{\bf n}^{\prime})$ in this difference we get finally the
$({\bf S}^{\prime}, {\bf n}^{\prime})$-expansion of the
solution $\phi^{i}_{[{\bf S},{\bf n}]} (\bm{\theta},X,T)$.

 The functions $({\bf S}^{\prime}, {\bf n}^{\prime})$
satisfy now the new deformed Whitham system 
(\ref{deform1})-(\ref{deform2}) corresponding to the
choice of the functions 
$\bm{\Phi}^{\prime}(\bm{\theta},{\bf k}^{\prime},
\bm{\omega}^{\prime}, {\bf n}^{\prime})$ as the functions
of zero approximation.

 Easy to see also that the transformation (\ref{sntrans})
remains polynomial in $X$-derivatives of 
$({\bf k},\bm{\omega},{\bf n})$ after the expression 
of time derivatives of $({\bf k},\bm{\omega},{\bf n})$
using the system (\ref{defwhsyst}).

 We obtain then that the transformation (\ref{sntrans})
gives a "triviality" connection between the systems
(\ref{defwhsyst}) written for the initial functions
$\bm{\Phi}(\bm{\theta}, \, {\bf k}, \, \bm{\omega}, \,
{\bf n})$ and
$\bm{\Phi}^{\prime}(\bm{\theta}, \, {\bf k}, \, 
\bm{\omega}, \, {\bf n})$ .

{\hfill Lemma 2.6 is proved.}

\vspace{0.5cm}

 At the end let us note again that the Lemma 2.6 is important
in fact for the integrable hierarchies rather than for the
one particular system (\ref{insyst}) according to
Dubrovin - Zhang approach to classification of integrable 
systems.

\vspace{0.5cm}

 I wish to thank Prof. B.A. Dubrovin, who suggested
the problem, for the interest to this work and many fruitful
discussions.

\end{document}